\documentclass[twocolumn,pre,superscriptaddress,notitlepage]{revtex4-1}

\PassOptionsToPackage{hyphens}{url}
\usepackage{hyperref}
\usepackage{url}
\usepackage{amsmath,amsfonts,amssymb}
\usepackage{epsfig}
\usepackage{subfigure}
\usepackage{graphicx}
\usepackage{textcomp}
\usepackage{float}
\usepackage[normalem]{ulem}
\usepackage{epstopdf}
\usepackage{subfigure}
\usepackage{dsfont}
\usepackage{color} %,soul
\usepackage[utf8x]{inputenc}
\usepackage{booktabs}

\newcommand{\hh}[1]{{\color{black} #1}}
\newcommand{\dmr}[1]{{\color{black} #1}}
\newcommand{\duncan}[1]{{\color{black} #1}}
\usepackage{fdsymbol}
\usepackage{graphicx} 
\begin{document}
\title{Examining the consumption of radical content on YouTube}

\author{Homa Hosseinmardi}
\affiliation{Department of Computer and Information Science, University of Pennsylvania, Philadelphia, PA 19104}
\author{Amir Ghasemian} 
\affiliation{Department of Statistics, Harvard University, Cambridge, MA 02138}
\affiliation{Department of Statistical Science, Fox School of Business, Temple University, Philadelphia, PA 19122}
\author{Aaron Clauset}
\affiliation{Department of Computer Science, University of Colorado Boulder, Boulder, CO 80309}
\affiliation{BioFrontiers Institute, University of Colorado Boulder, Boulder, CO 80303}
\affiliation{Santa Fe Institute, Santa Fe, NM 87501}
\author{Markus Mobius}
\affiliation{Microsoft Research New England, Cambridge, MA 02142}
\author{David M. Rothschild}
\affiliation{Microsoft Research New York, New York, NY 10012}
\author{Duncan J. Watts}
\affiliation{University of Pennsylvania, Philadelphia, PA 19104}
\affiliation{The Annenberg School of Communication, University of Pennsylvania, Philadelphia, PA 19104}
\affiliation{Information, and Decisions Department, University of Pennsylvania, Philadelphia, PA 19104}
 

\begin{abstract}
Although it is understudied relative to other social media platforms, YouTube is arguably the largest and most engaging online media consumption platform in the world. Recently, YouTube's scale has fueled concerns that YouTube users are being radicalized via a combination of biased recommendations and ostensibly apolitical ``anti-woke'' channels, both of which have been claimed to direct attention to radical political content. Here we test this hypothesis using a representative panel of more than 300,000 Americans and their individual-level browsing behavior, on and off YouTube, from January 2016 through December 2019. Using a labeled set of political news channels, we find that news consumption on YouTube is dominated by mainstream and largely centrist sources. Consumers of far-right content, while more engaged than average, represent a small and stable percentage of news consumers. However, consumption of ``anti-woke'' content, defined in terms of its  opposition to progressive intellectual and political agendas, grew steadily in popularity and is correlated with consumption of far-right content off-platform. 
%Finally, we find no evidence that engagement with either far-right or reactionary content is caused by YouTube recommendations \hh{systematically}. 
We find no evidence that engagement with far-right content is caused by YouTube recommendations systematically, nor do we find clear evidence that anti-woke channels serve as a gateway to the far right. Rather, consumption of political content on YouTube appears to reflect individual preferences that extend across the web as a whole. 
%Our results emphasize the importance of measuring consumption directly, \dmr{recognizing reader demand, }rather than inferring consumption from recommendations.  

\end{abstract}

\maketitle

As affective political polarization rises in the US~\cite{iyengar2019origins} and trust in traditional sources of authority declines~\cite{jones2015declining,tucker2018social}, concerns have arisen regarding the presence, prevalence, and impact of false,  hyperpartisan, or conspiratorial content on social media platforms.
Most research on the potentially polarizing and misleading effects of social media has focused on Facebook and Twitter~\cite{conover2011political,del2016echo, grover2019polarization,bossetta2018digital,alizadeh2020content,aral2019protecting, lazer2018science,pennycook2019fighting,roozenbeek2019fake}, reflecting a common view that these platforms are the most ``news-oriented'' social media platforms. However, 
roughly $23$ million Americans rely on YouTube as a source of news~\cite{konitzer2020measuring,emarketer_youtube}, a number comparable to the corresponding Twitter audience~\cite{konitzer2020measuring,businessofapps_twitter}, and is growing both in size and engagement. YouTube news content spans the political spectrum, and includes content producers of all sizes. Recent work~\cite{clark2020understanding} has identified a large number of YouTube channels, mostly operated by individuals or small organizations, that promote a collection of ``far-right'' ideologies (e.g. white identitarian) and conspiracy theories (e.g. QAnon).  The popularity of some of these channels, along with salient popular anecdotes, have prompted claims that YouTube's recommendation engine \hh{systematically} drives users to this content, and effectively radicalizing its users~\cite{rabbit_hole, youtube_radical, wrong_internet, whyRight}. For example, it has been reported that starting from factual videos about the flu vaccine, the recommender system can lead users to anti-vaccination conspiracy videos \cite{youtube_radical}.

Recent qualitative work~\cite{lewis2018alternative} has identified a separate collection of channels labeled variously as ``Reactionary,'' ``Anti woke'' (AW), ``Anti Social Justice Warriors''(ASJW), ``Intellectual Dark Web'' (IDW), or simply ``anti-establishment.'' Although these channels do not identify themselves as politically conservative, and often position themselves as non-ideological or even liberal ``free thinkers,'' in practice their positions are largely defined in opposition to progressive social justice movements, especially those concerning identity and race, as well as critiquing institutions such as academia and mainstream media for their ``left-wing bias''~\cite{lewis2018alternative,lewis2020news}. Concurrently, ``anti-woke'' rhetoric has increasingly been adopted by mainstream \dmr{Republican} politicians~\cite{bacon_anti_woke}, undermining claims that it is intrinsically apolitical. While anti-woke YouTube channels typically do not explicitly endorse far-right ideologies, some channel owners invite guests who are affiliated with the far right onto their shows and allow them to air their views relatively unchallenged, thereby effectively broadcasting and legitimizing far-right ideologies~\cite{lewis2018alternative}. If these channels act as a kind of gateway to the far right, they would constitute a related-yet-distinct radicalization mechanism from the recommendation system per-se\hh{~\cite{rabbit_hole,ribeiro2019auditing}}. Based on these considerations, and recognizing that any label for this loose collection of channels is likely to be inaccurate for at least some members, we refer to them hereafter as anti woke (AW). 

%While anti-woke channel owners do not typically endorse far-right ideologies explicitly, and in some cases directly oppose them, some channel owners have been accused of…

%\hl{Although reports of various mechanisms driving people to far-right videos raise legitimate concerns, quantitative evidence to support them has proven elusive.} 

Although reports of various mechanisms driving people to politically radical content has received great attention, quantitative evidence to support them has proven elusive. On a platform with almost 2 billion users~\cite{youtub_stats}, it is possible to find examples of almost any type of behavior; hence anecdotes of radicalized individuals~\hh{\cite{rabbit_hole}}, however vivid, do not on their own indicate systematic problems. Thus, the observation that a particular mechanism (e.g. recommendation systems steering users to extreme content; far-right personalities appearing on anti-woke channels acting as gateways to the far right) might plausibly have a large and measurable effect on audiences does not substitute for measuring the effect.  
Finally, the few empirical studies
\cite{ribeiro2019auditing,cho2020search,munger2020right,ledwich2019algorithmic,faddoul2020longitudinal} that have examined the question of YouTube radicalization have reached conflicting conclusions, with some 
finding evidence for it~\cite{ribeiro2019auditing,cho2020search} and others finding the opposite~\cite{munger2020right,ledwich2019algorithmic}. These disagreements may arise from methodological differences that make results difficult to fairly compare---for example, Ref. \cite{ledwich2019algorithmic} examines potential biases in the recommender by simulating logged-out users, whereas Ref. \cite{ribeiro2019auditing} reconstructs user histories from scraped comments. The disagreement may also reflect limitations in the available data, which is intrinsically ill-suited to measuring either individual or aggregate consumption of different types of content over extended time intervals, such as user sessions or ``lifetimes.'' Absent such data for a large, representative sample of real YouTube users, it is difficult to evaluate how much far-right content is in fact being consumed (vs. produced), how it is changing over time, and to what extent it is being driven by YouTube's own recommendations, spillovers from anti-woke channels, or other entry points.   
%\hh{editor asked more general interest motivation here, similar to example of conspiracy}
%Political polarization and disinformation both can potentially harm democracy, and accent each other~\cite{tucker2018social}. YouTube is of special interest as an understudied corner of internet, with without the usual editorial filter in traditional media, can host conspiracy stories, and potential echo chambers \cite{marwick2017media}. 

Here we investigate the consumption of radical \hh{political} news content on YouTube using a unique data set comprising a large ($N=309,813$) representative sample of the US population, and their online browsing histories, both on and off the YouTube platform, spanning four years from Jan 2016 to Dec 2019. To summarize, we present five main findings.  (i) Consistent with previous estimates~\cite{allen2020evaluating}, we find that total consumption of any news-related content on YouTube accounts for $11\%$ of overall consumption and is dominated by mainstream, and generally centrist or left-leaning, sources. 
(ii) The consumption of far-right content is small both in terms of number of viewers and total watch time, where the former decreased slightly and the latter increased slightly over the observation period. %\dmr{Despite tracking meaningful production, the consumption of far-left content is negligible}
(iii) In contrast, the consumption of anti-woke content, while also small relative to mainstream or left-leaning content, grew both in numbers of users and total watch time. (iv) The pathways by which users reach far-right videos are diverse and only a fraction can plausibly be attributed to platform recommendations.
Within sessions of consecutive video viewership, we find no trend toward more extreme content, either left or right, indicating that consumption of this content is determined more by user preferences than by recommendation. (v) Consumers of anti-woke, right, and far-right content also consume a meaningful amount of far-right content elsewhere online, indicating rather than the platform \dmr{(either the recommendation engine or consumption of anti-woke content)} pushing them towards far-right content, it is compliment to their larger news diet.
%\hh{It is worth mentioning that the far-left community is very small both in size and consumption, and this imbalance resulted in most of our conclusions in regard to radical political content to discuss the far-right category}.

% editor's comment: has value judgments and needs revision 
%\amir{We conclude that while previously expressed concerns about the rise of anti-woke commentary on YouTube appear to be supported by our data, to some extent, the focus on the recommendation engine is overly narrow.}
These results indicate little evidence for the popular claim that YouTube drives users to consume more radical political content, either left or right. Instead we find strong evidence that, while somewhat unique with its growing and dedicated anti-woke channels, YouTube should otherwise be viewed as part of a larger information ecosystem in which conspiracy theories, misinformation, and hyperpartisan content 
are widely available, easily discovered, and actively sought out~\cite{munger2020right,wilson2020cross}.

\section*{Methods and Materials}
 Our data are drawn from Nielsen’s nationally representative desktop web panel, spanning January 2016 through December 2019, which records individuals' visits to specific URLs. We use the subset of $N =309,813$ panelists who have at least one recorded YouTube pageview. Parsing the recorded URLs, we found a total of $21,385,962$ watched-video pageviews (Table~\ref{basic_stats}). %We then annotated 
We quantify the user's attention by the duration of in-focus visit to each video in total minutes~\cite{lazer2020studying}. Duration or time spent is credited to an in-focus page and when a user returns to a tab with previously loaded content, duration is credited immediately. Each YouTube video has a unique identifier embedded in its URL,  yielding $9,863,964$ unique video IDs. To post a video on YouTube, a user must create a channel with a unique name and channel ID. For all unique video IDs, we used the YouTube API to retrieve the corresponding channel ID, as well as metadata such as the video's category, title, and duration. We then labeled each video based on the political leaning of its channel. 

%%%%%%%%%%%%%%%%%%%%
%% TABLE 1
\begin{table}[b!]%[tbhp]
\centering
\caption{YouTube data descriptive statistics.}
\begin{tabular}{lr}
\hline
%\midrule
 Number of unique users & $309,813$  \\ \hline
 Number of watched-video pageviews & $21,385,962$ \\ \hline
 Number of unique video IDs & $9,863,964$ \\ \hline
 Number of unique channel IDs & $2,293,760$  \\ \hline
 Number of sessions & $8,620,394$ \\ \hline
%\bottomrule
\end{tabular}\label{basic_stats}
\end{table}
%%%%%%%%%%%%%%%%%%%%

\paragraph{Video labeling.} Previous studies~\cite{lewis2018alternative,ribeiro2019auditing,ledwich2019algorithmic,munger2020right,faddoul2020longitudinal} have devoted considerable effort to labeling YouTube channels and videos based on their political \hh{content}. In order to maintain consistency with the existing research literature, we derived our labels from two of these previous studies which collectively classified over 1,100 YouTube channels. First, Ref.~\cite{ledwich2019algorithmic} classified 816 channels along a traditional left/center/right ideological spectrum as well as a more granular categorization into 18 tags such as Socialist, Anti-Social Justice Warriors (ASJW), Religious Conservative, White Identitarian, and Conspiracy. Second, Ref.~\cite{ribeiro2019auditing} classified 281 channels as belonging to one of Intellectual Dark Web (IDW), Alt-lite, or Alt-right, and a set of 68 popular media channels as Left, Left-Center, Center, Right-Center, or Right. 
Because the two sources used slightly different classification schemes, we mapped their labels to a single set of six categories: far left (fL), left (L), center (C), anti-woke (AW), right (R), and far right (fR). Five of the six categories (fL,~L,~C,~R,~fR) fall along a conventional left-right ideological spectrum. For example, YouTube videos belonging to channels with ideological labels such as ``Socialist'' are considered further to the left of ``left'' content and hence were assigned to far left, whereas ``Alt-right'' is a set of ideologies that exemplify extreme right content and so were assigned to far right \cite{ribeiro2019auditing,munger2020right}. The ``Anti Woke'' (AW) category mostly comprises the labels ``Intellectual Dark Web'' (IDW)~\cite{ribeiro2019auditing} and ``Anti Social Justice Warriors'' (ASJW)~\cite{ledwich2019algorithmic}, but also a small set of channels labeled ``Men's Right Activists'' (MRA)~\cite{ribeiro2019auditing,ledwich2019algorithmic}, is more easily defined by what it opposes---namely progressive social justice movements and mainstream left-leaning institutions---than what coherent political ideology it supports~\cite{lewis2018alternative}. 
\duncan{Reflecting its non-traditional composition, we locate it to the right of center but left of right and far right. We note that the ordering of the categories from left to right, while helpful for visualizing results in some cases, is not important to any of main findings.} 
%we resist placing it on a conventional L-R spectrum, instead representing it separately (although, as we will show later, it exhibits much stronger associations with right and far-right categories than with left or centrist).
Overall, our data cover 974 channels, corresponding to $523,242$ videos (following ~\cite{ribeiro2019auditing} and ~\cite{ledwich2019algorithmic} all videos published by a channel under study received the channel's label). % a sample that accounts for around 35\% of YouTube's total news consumption (SI Appendix, Fig. %\ref{fig:news_share}S2).
%Details on the number of videos in each category, the assignment of channels and references used can be found in SI Appendix, Tables S1 and S2, section C.

\paragraph{Label imputation.}Using the YouTube API, $20.1\%$ of the video IDs had no return from the API (we refer to these as unavailable videos), a problem that previous studies also faced~\cite{wu2018beyond}. The YouTube API does not provide any information about the reason for this return value; however, for some of these videos, the YouTube website itself shows a ``sub-reason'' for the unavailability. We crawled a uniformly random set of $368,754$ videos and extracted these sub-reasons from the source HTML. Stated reasons varied from video privacy settings to video deletion or account termination as a result of violation of YouTube policies. For channels such as ``infowars,'' which was terminated for violating YouTube’s Community Guidelines~\cite{terminated_infowars}, none of their previously uploaded videos are available through the YouTube API. Therefore, it is important to estimate the fraction of these unavailable videos that will receive each of the political leaning labels, and whether that distribution will affect our findings. 

To resolve this ambiguity, we treated it as a missing value problem and imputed missing video labels via a supervised learning approach. To obtain accurate labels for training such a model, we first searched for the overlap between our set of unavailable videos and data sets from previous studies~\cite{faddoul2020longitudinal,ribeiro2019auditing,munger2020right,wu2018beyond,ledwich2019algorithmic}, which had collected the metadata of many now unavailable videos at a time when they still existed on the platform. 
%(SI Appendix, Table %\ref{table:imputation_gt}
%S4). 
This approach yielded channel IDs for $69,611$ of our unavailable videos. We then trained a series of classifiers, which we used to impute the labels of the remaining unavailable videos. For the features of the supervised model, we extracted information surrounding each unavailable video, such as the web partisan score of news domains viewed by users before and after it, along with the YouTube and political categories of all videos watched in close proximity within the same session. We also exploited a set of user-level features, such as the individual's monthly consumption from different video and web categories during their lifetime in our data. %Details on the feature engineering and model selection can be found in SI Appendix, section D. 
 
For each political channel category we trained a binary random forest classifier over 96 predictors, which yielded an AUC (Area Under Curve) of $0.95$ for far left, $0.98$ for left, $0.95$ for center, and $0.97$ for anti woke, right, and  far right, on the holdout set. To assign labels, we consider two different thresholds, one with high precision and one with high recall.% (SI Appendix, Tables %\ref{table:video_type_imputed}
%S5-S6). 
 ~For all presented results in the main text, we use a high precision imputation model, which means fewer videos (with higher confidence) are retrieved from the positive class. Therefore, the results presented here reflect the lower bound of the number of videos in each political category. As a robustness check, we repeat some of our experiments using the upper bound.% (high recall; see SI Appendix, section I).

\paragraph{Constructing sessions.}Previous studies have analysed web browsing dynamics by breaking a sequence of pageviews into sub-sequences called sessions~\cite{kumar2010characterization}. In this work, we define a YouTube session as a set of near-consecutive YouTube pageviews by a user. 
Within a YouTube session, a gap less than $\delta$ minutes is allowed between a YouTube non-video URL and the next YouTube URL, or a gap less than $\gamma$ minutes is allowed between a YouTube video URL and the next YouTube URL, otherwise the session breaks and a new session will start with the next YouTube URL.
External pageviews (all non-YouTube URLs) are allowed within these gaps. For brevity, throughout the rest of the paper, we refer to a YouTube session as simply a session. In the main text we will present the results for sessions created by $\delta=10$  and $\gamma=60$ minutes. To check the robustness of our findings to these choices, we repeated the session-level experiments with different values of $\delta$ and $\gamma$.% (see SI Appendix, section I).

%%%%%%%%%%%%%%%%%%%%
%% FIGURE 1
\begin{figure}[tb!]
\centering
\includegraphics[width=9cm]{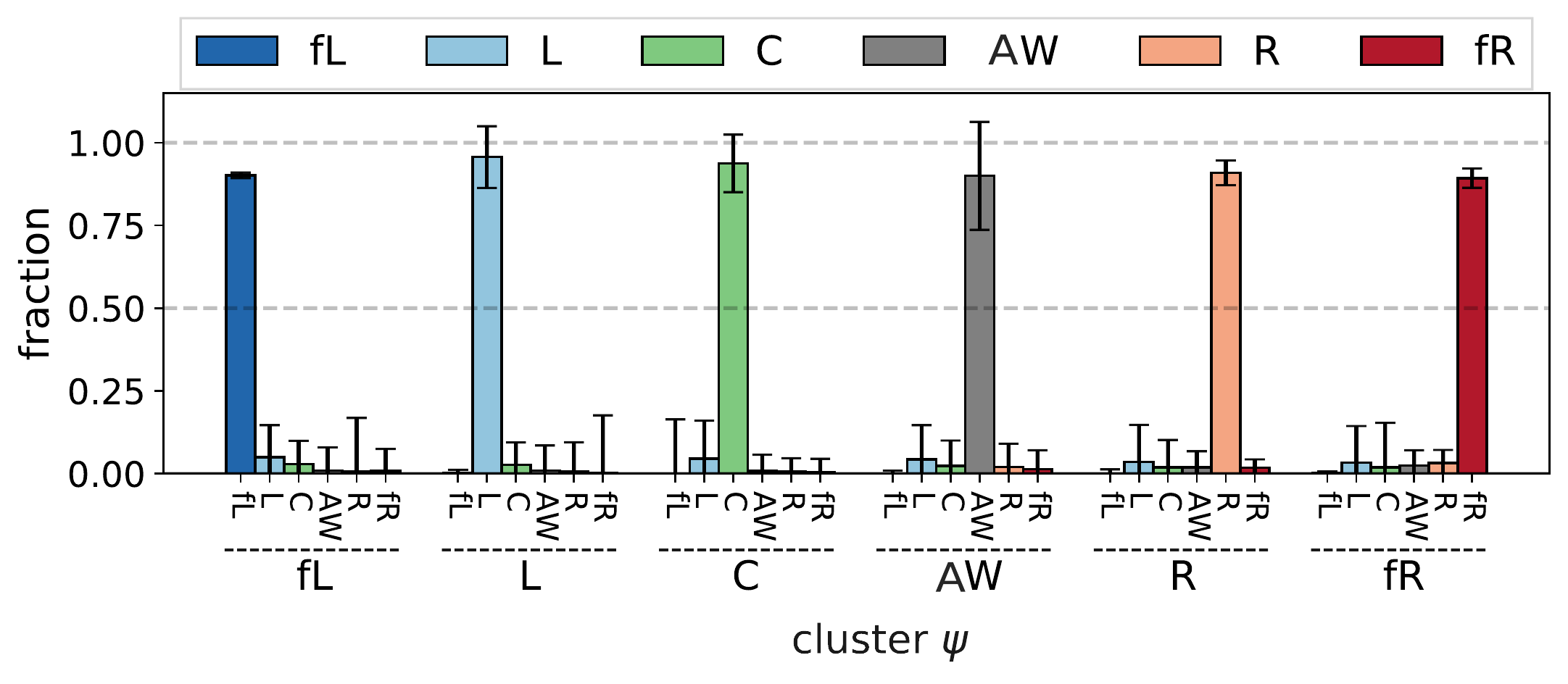}
\caption{The archetypes of news consumption behavior on YouTube for each cluster.}
\label{fig:archetypes}
\end{figure}
%%%%%%%%%%%%%%%%%%%%

\paragraph{User clustering.} An individual is considered a news consumer if,  over the course of one month, they spend a minimum of 1 minute watch-time on any of the political channels in our labeled set.
Each month we characterized every individual who consumed news on YouTube in terms of their normalized monthly viewership vector $\nu^m_i$ whose $j$-th entry, $\nu^m_{ij}$ corresponds to the fraction of viewership of user $i$ from channel category $j$ ($j \in \{{\rm{fL,~L,~C,~AW,~R,~fR}}\}$). 
We then used hierarchical clustering to assign each individual to one of $K$ communities of similar YouTube news diets, with $K$ in the range from 2 to 12. Running 19 different measures of model fit to find the optimal number of communities, 6 and 5 had equally the highest number of votes~\cite{NbClust}, where we set the number of communities to be $K=6$ to capture all categories.%Using the silhouette method~\cite{de2015recovering}, we found the optimal number of communities to be $K=5$ (SI Appendix, Fig.%\ref{fig:silhouette_clustering}X).
~For each of these six clusters we then identified its centroid obtained by averaging the normalized  monthly viewership vectors of all cluster members.% (see SI Appendix, section E for details). 
~Finally, we labeled each community as $\psi(t)$ ($\psi(t)\in\{{\rm{fL, L, C, AW, R, fR}}\}$) according to the predominant content category of its centroid. %As a robustness check, we performed similar analysis with nonzero news consumption and at least 10 and 30 minutes news consumption per month as more relaxed and more strict definitions of ``news consumers'' (see SI Appendix, Figs. %\ref{fig:cluster_vs_threshold}S23).
As a robustness check, we performed similar analysis with more relaxed and more strict definitions of ``news consumers.''% (see SI Appendix, Section I).

\section*{Results}
Before we examine the degree to which YouTube’s recommendation engine drives users to more extreme forms of political content, we first present a series of analyses to characterize and quantify the overall consumption patterns among different types of content. These patterns allow us to test for several confounding possibilities in the dynamics of news consumption on YouTube, and provide a clear background picture against which to measure systematic deviations caused by recommendations.
 
These six clusters correspond closely with our six categories of political content (Fig. \ref{fig:archetypes}): the centroid or ``archetype member'' of each cluster devotes more than 90\% of their attention to just one content category, with the remaining 10\% distributed roughly evenly among the other categories. In fact, the consumption patterns of individual users in each of these clusters aligns strongly with their containing cluster, so much so that more than 70\% receive at least 80\% of their content from one content type and more than 95\% receive at least 50\% of their content from one content type.%, SI Appendix, Fig. S3.
~This result has two important implications for our analysis. \duncan{First, it reveals that YouTube users are meaningfully associated with definable ``communities'' in the sense that consumption preferences are relatively homogeneous within each community and relatively distinct between them. Throughout the remainder of the paper, we will use the term community and cluster interchangeably, and ``category'' to refer to the corresponding population of videos.}
%These ``information community'', can be defined as a set of individuals who are exposed to or primarily consume a single ideologically consistent type of information
% Because these communities strongly resemble so-called ``echo chambers''~\cite{sunstein1999law}, in which individuals are exposed to ideologically consistent information, we will use these terms interchangeably throughout the remainder of the paper. 
%Second, it demonstrates that the political content categories we have identified capture a large fraction of observed behavior in a parsimonious way \hh{(see SI appendix as a robustness check for more restrict definitions of news consumers, Fig. S18-S19)}. We present our main results in terms of these communities. 
Second, it demonstrates that the political content categories we used align closely with the actual behavior of users in a parsimonious way.% (see SI Appendix as a robustness check for more strict definitions of news consumers, Fig. S3-S4). 
~We present our main results in terms of these communities. 
%%%%%%%%%%%%%%%%%%%%
%% FIGURE 2
\begin{figure}[tb!]
\centering
\includegraphics[width=9cm]{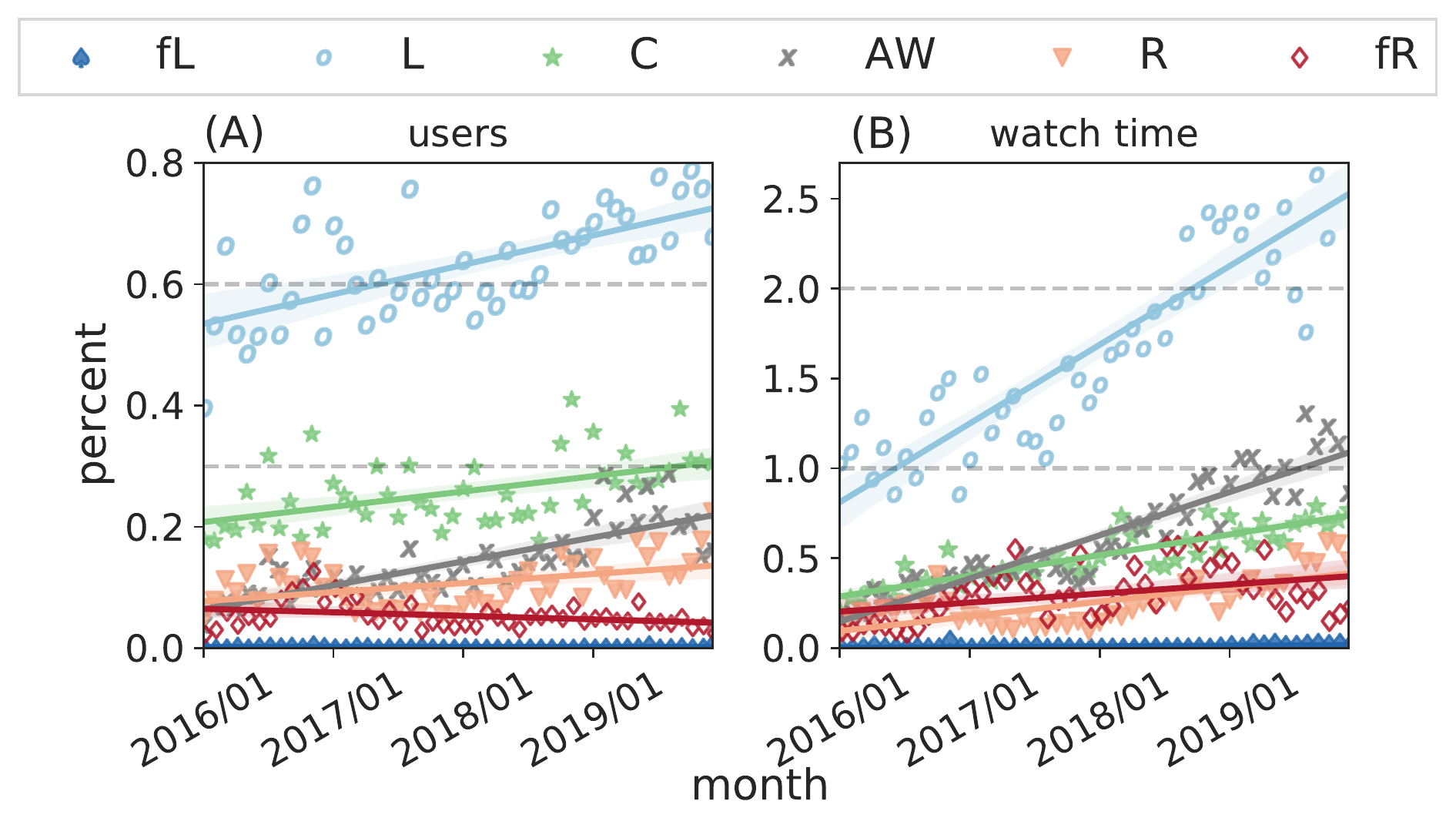}
\caption{Breakdown of percent of (A) users, (B) consumption falling into the six political channel categories, per month, January $2016$ to December $2019$. Panel (A) is the percent of users falling into each community, and panel (B) presents the percentage of viewership duration from each channel category.}
\label{fig:trends}
\end{figure}
%%%%%%%%%%%%%%%%%%%%

\paragraph{Community engagement.} 

To check for any overall trends \duncan{in category preferences, we examine changes in total consumption associated with each of the six communities over the four-year period of our data,} quantified both in terms of population size (Fig.~\ref{fig:trends}A) and total time spent watching (Fig.~\ref{fig:trends}B).% (see SI Appendix, Fig. S5 %\ref{fig:session_view}
%for two other related metrics: page view counts and session counts, and Tables S8-S9 for more details on fitted lines.).  
\duncan{
~Consistent with previous work~\cite{allen2020evaluating}, we find that total news consumption accounts for only $11\%$ of total consumption.% (see SI Appendix, section D for definition of ``news''). 
~Of this total, the 974 channels for which we have political labels account for roughly one third (i.e. $3.32\%$ of total watch time).}
%accounted for only a small fraction of overall YouTube activity, whether in terms of population of active news consumers \hh{($1.19\%$)} or watch time ($3.32\%$) averaged over the four year period. (Reported watch time is lower than the $11\%$ which included all channels labeled by YouTube as ``news'' whereas these analyses were restricted to the 974 channels for which we had political labels.) 
Moreover, the largest community of news consumers---both in terms of population size ($0.63\%$) and watch time ($1.65\%$)---was the ``left'' mainstream community. Fig.~\ref{fig:trends}A shows that the far-right community was the second smallest (after far-left) by population ($0.05\%$) and declined slightly in size after a peak at the end of 2016. In contrast, the anti-woke category started roughly at the same size but grew considerably, overtaking right and almost matching center. Next, %Fig.~\ref{fig:trends}B shows that watch time for both far-right and anti-woke news grew over our observation period, where the former increased two-fold from 0.2\% of total consumption in Jan 2016 to 0.4\% in Dec 2019 and the latter grew even more rapidly from 0.13\% to 1.09\%. 
%Fig.~\ref{fig:trends}B shows that watch time for both far-right and AW news grew (slope = 0.004 percent per month, p-value = 0.004 and slope = 0.021 percent per month, p-value $< 10^{-16}$ respectively), where the latter grew more rapidly from 0.16\% of total consumption in Jan 2016 to about 0.8\% in Dec 2019 . 
Fig.~\ref{fig:trends}B shows that monthly watch time for both far-right and anti-woke news grew over our observation period, where the former increased almost two-fold from an average of $0.17\%$ of total consumption in $2016$ to $0.30\%$ in $2019$ ($t=-3.17,~P < 0.005$) and the latter grew more rapidly from an average of $0.31\%$ to $1.02\%$ ($t=-14.08,~P < 10^{-5}$).
In both cases, watch time grew faster than the overall rate of growth of news consumption (8\% of average monthly consumption in 2016 to 11\% in 2019 ($t=-13.13,~P < 10^{-5}$)) and the anti-woke community ended the period accounting for more watch time than any category except left. \duncan{We note that these data exhibit strong variations, and hence we caution against extrapolating these trends into the future.} % the observed trends may not hold up out of sample
Because the far-left community is the smallest by population size ($0.002\%$) and watch time ($0.009\%$) (being barely visible on the scale of Fig~\ref{fig:trends}), we drop it from our results if the low observation count makes statistical inferences unreliable.

%%%%%%%%%%%%%%%%%%%%
%% FIGURE 3
\begin{figure}[tb!]
\centering
\includegraphics[width=7.7cm]{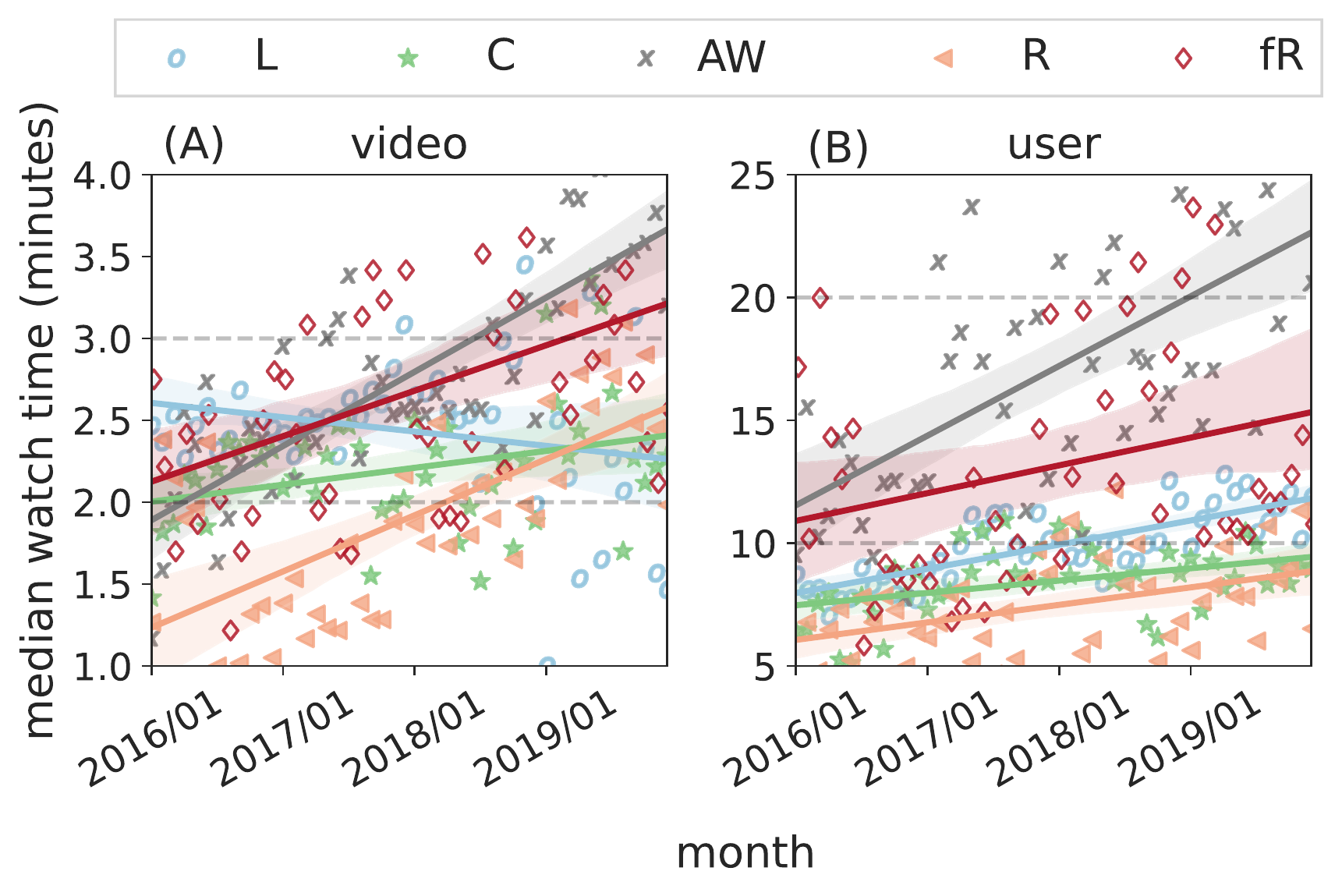}
\caption{(A) Median monthly video consumption (minutes) across different channel types, and (B) median user consumption (minutes) within each community.}
\label{fig:consumption_per_unit}
\end{figure}
%%%%%%%%%%%%%%%%%%%%

%These results indicate that the phenomenon of right-wing radical and reactionary content on YouTube, while small in relative terms, nonetheless affects more than half a million Americans monthly, averaged over the four year period, and growing both in population size and on consumption.
\duncan{These results indicate that news in general and far-right content in particular, account for only a small portion of consumption on YouTube. As emphasized previously, however, even a small percentage of users or consumption time can translate to large absolute numbers at the scale of YouTube. 
%the phenomenon of right-wing radical and reactionary content on YouTube while small in relative terms, nonetheless affects 
For example, averaged over the four year period, $335,209$ Americans consumed far-right content and $764,405$ consumed anti-woke content at least once in a given month, where the latter grew steadily in population and watch time.} This result is also robust to other choices of consumption metric (e.g. page views or session counts), threshold for inclusion in the ``news consuming population,'' and imputation model.% (SI Appendix, Figs. S5, S13, and S16-17% \ref{fig:session_view},  \ref{fig:cluster_vs_threshold}, \ref{fig:cluster_bounds} and \ref{fig:trends_S})
%). 
To better understand these dynamics we now investigate individual-level behavior.

\paragraph{Individual engagement.} Complementing the aggregate (community-level) results in ~Fig.~\ref{fig:trends}, Fig.~\ref{fig:consumption_per_unit} shows both absolute levels of, and changes over time in, consumption measured at the individual level of videos (Fig.~\ref{fig:consumption_per_unit}A) and users (Fig.~\ref{fig:consumption_per_unit}B), respectively.% (See SI Appendix, Tables S10-S11 for more details on fitted lines). %Fig.~\ref{fig:consumption_per_unit} presents further evidence of the appeal of radical and reactionary content. 
\duncan{~Fig.~\ref{fig:consumption_per_unit} presents evidence of stronger engagement with far-right and anti-woke content than for other categories. %(SI Appendix, Figs. S6-S8 and Tables S12-S13).
The median per-video watch time of far-right videos increased by 50\% from an average of 2 minutes in 2016 to 3 minutes in 2019 ($t=-3.43, P < 0.005$), while the per-video watch time of anti-woke category roughly doubled ($t=-8.55, ~P < 10^{-5}$) over the same period, starting out well below centrist and left-leaning videos but eventually overtaking all of them. 
% Second, the median per-month watch time for individual members of the far-right and AW communities also increased by more than 150\% over the same time period, from 10 minutes per month to roughly 15 minutes and to more than 20 minutes per month respectively, almost twice engagement as users in left, center and right communities.
Second, the median per-month watch time for individual members of the far-right and anti-woke communities is up to almost twice the engagement of users in left, center and right communities ($P~< 10^{-4}$; for all pairwise comparisons of far right and anti woke with left, center, and right). In other words, while far-right and anti-woke communities remained relatively small in size throughout the observation period (with only anti-woke growing), their user engagement grew to exceed that of every other category% (see SI Appendix as a robustness check for more strict definitions of news consumers, Fig. S14)
.}   

%%%%%%%%%%%%%%%%%%%%
%% FIGURE 4
\begin{figure}[tb!]
\centering
\includegraphics[width=6.5cm]{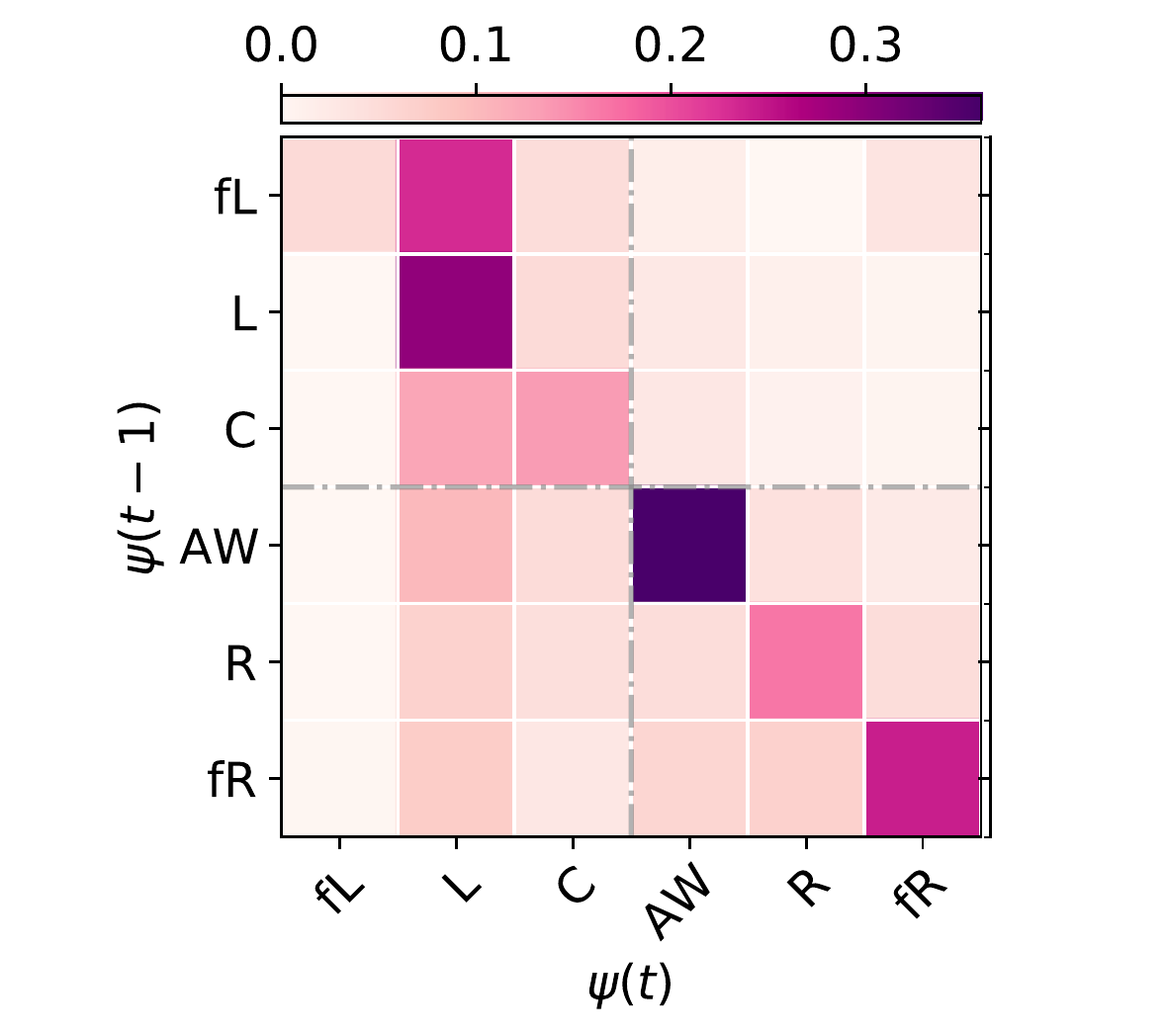}
\vspace{-2mm}
\caption{A heatmap showing the probability that an individual from cluster $\psi(t-1)$ at month $t-1$ will move to cluster $\psi(t)$ at month $t$. Each month, users may not fall into any of these communities, if they are not among ``news consumers'' in that particular month. %The bar plot presents the steady states of the transition matrix, as the fraction of YouTube users falling into each cluster, when $t \to  \infty $.
}
\label{fig:clustering}
\end{figure}
%%%%%%%%%%%%%%%%%%%%

Further examining individual behavior, Fig. \ref{fig:clustering} shows the average probability $P(\psi_j(t)|\psi_i(t-1))$ of an individual member of community $\psi_i$ in month $t-1$ moving to community $\psi_j$ in month $t$. 
As indicated by darker shades along the diagonal, the dominant behavior is for community members to remain in their communities from month to month, suggesting that all communities exhibit ``stickiness.'' Moreover, when individuals do switch communities they are more likely to move from the right side of the political spectrum to the left  than the reverse, while individuals in the center are more likely to move left than right. 
Also, there is more between-community movement from right and far-right than left and center to the anti-woke community (i.e.  $P(\psi(t)={\rm{AW}}|\psi(t-1)={\rm{fR}})=0.06,~P(\psi(t)={\rm{AW}}|\psi(t-1)={\rm{R}})=0.05$ vs. $P(\psi(t)={\rm{AW}}|\psi(t-1)={\rm{C}})=0.03,~P(\psi(t)={\rm{AW}}|\psi(t-1)={\rm{L}})=0.02$ and $P(\psi(t)={\rm{AW}}|\psi(t-1)={\rm{fL}})=0.01$), indicating that the anti-woke community gains more audience from right-wing than from center and left-wing. Also, the most common transition to far right is from right, far left, and anti woke (i.e. $P(\psi(t)={\rm{fR}}|\psi(t-1)={\rm{R}})=0.04>P(\psi(t)={\rm{fR}}|\psi(t-1)={\rm{fL}})=0.03>P(\psi(t)={\rm{fR}}|\psi(t-1)={\rm{AW}})=0.02>P(\psi(t)={\rm{fR}}|\psi(t-1)={\rm{C}})=0.01>P(\psi(t)={\rm{fR}}|\psi(t-1)={\rm{L}})=0.00$).

%Finally, the gray horizontal bars in Fig. \ref{fig:clustering} show the steady states of the transition matrix, representing the fraction of YouTube users who would hypothetically reside in each community in the limit $t \to  \infty $ (assuming constant transition probabilities). In this simulated equilibrium, the left and center communities are  projected to remain the largest groups.

%Finally, the gray horizontal bars in Fig. \ref{fig:clustering} show the steady states of the transition matrix, representing the fraction of YouTube users who would hypothetically reside in each community in the limit $t \to  \infty $ (assuming constant transition probabilities). In this simulated equilibrium, the left and center communities are  projected to remain the largest groups with $0.76\%$ and $0.29\%$ of the population respectively. The far-right community is projected to three times its size, growing to $0.84\%$ of YouTube users from $0.22\%$ in Dec 2019. 

%%%%%%%%%%%%%%%%%%%%
%% FIGURE 5
\begin{figure}[tb!]
\centering
\includegraphics[width=8cm]{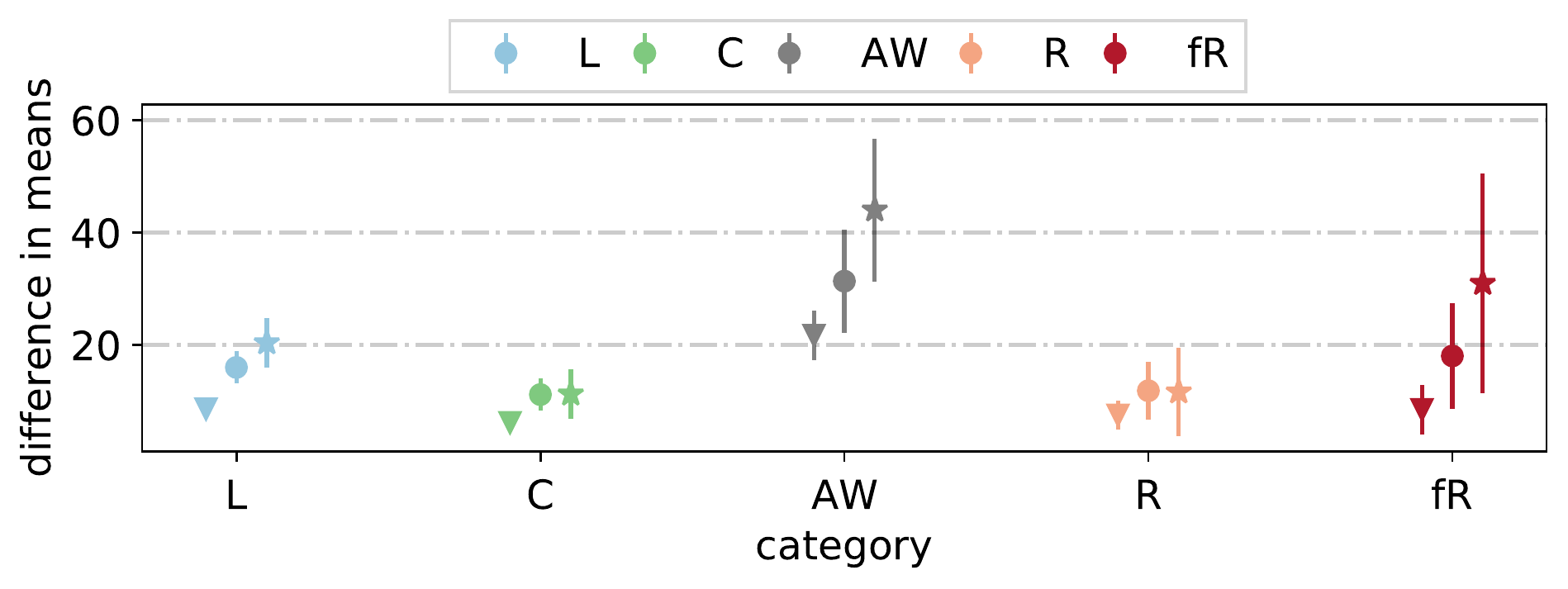}
\caption{Difference in means of daily consumption change, in the event of bursty consumption from a specific political category. Individuals are assigned either to bursty consumption group in the event of watching at least $M^k_v$ videos from category $k$ ($k\in\{{\rm{L, C, AW, R, fR}}\}$) within a session, or in control group, if none of their sessions has more than one video from same category with at least $M^k_v$ videos in their lifetime.
We run three experiments with different values of $M^k_v$, where $\blacktriangledown:~M^k_v=2,~  \medblackcircle:~M^k_v=3,~\medblackstar:~M^k_v=4$. Markers show the difference in means, and the vertical lines present the 95\% confidence interval.
The exposure can be driven by user, recommendation or external sources. Difference in change of daily consumption, after bursty consumption, is almost twice bigger for AW compared to the other political categories, when controlled for other covariates. }
\label{fig:burstiness}
\end{figure}
%%%%%%%%%%%%%%%%%%%%

\paragraph{Concentrated exposure predicts future consumption.}
Exposure to concentrated ``bursts'' of radical content may correlate with future consumption more strongly than equivalent exposure to other categories of content~\cite{Bail9216}. To check for this possibility, we define a ``burst'' of exposure as consumption of at least $M^{(k)}_v$ videos of category $k$ ($k\in\{{\rm{L, C, R, AW, fR}}\}$) within a single session, and a ``treatment event'' as the first instance in a user's lifetime when they are exposed to such a burst. We dropped the far-left category as the number of samples were too small for this experiment.

For each content category, we consider three ``treatment groups'' comprising individuals who are exposed to burst lengths $M^{(k)}_v\in\{2,3,4\}$ %(SI Appendix, Table S14% \ref{table:number_user_burtiness}) 
and compute the difference in their average daily consumption of the same content category $k$ pre- and post-exposure. Finally, we compute the difference in difference between our treatment groups and a ``control group'' of individuals with maximum $M^{(k)}_v=1$ video per session, where we use propensity score matching to account for differences in historical web and YouTube consumption rates, demographics (age, gender, race), education, occupation, income, and political leaning.% (see SI Appendix, section F for details). 
~We emphasize that these treatments are not randomized and could be endogenous (i.e. individuals are exposed to longer bursts because they already have higher interest in the content), hence the effects we observe should not be interpreted as causal. As a purely predictive exercise, this analysis reveals whether exposure to a fixed-length burst of content at one point in time has different effects at a future point in time across content categories.

Fig. \ref{fig:burstiness} shows the results for $M^{(k)}_v =\{2,~3,~4\}$ across our five remaining content categories. In all cases, 
increases in burst length from 2 to 4 correspond to higher future consumption relative to the control groups. Individuals exposed to bursts of anti-woke content show much larger effects than other content categories for all burst lengths, and larger marginal effects for longer vs. shorter bursts. 
The daily far-right content consumption of individuals exposed to far-right bursts of length  $M^{{\rm{(fR)}}}_v = 4$, increases by a gap of almost 30 seconds post-exposure, relative to the control group ($\tau=30.99 \pm 19.50$ with CI=$0.95$).

\paragraph{Potential causes of radicalization.} Summarizing thus far, consumption of far-right and anti-woke content on YouTube---while small relative to politically moderate and non-political content---is stickier and more engaging than other content categories; and, in the case of anti woke, is increasingly popular. Previous authors have argued that the rise of radical content on YouTube is somehow driven by the platform itself, in particular by its built-in recommendation engine~\cite{youtube_radical,rabbit_hole}. While this hypothesis is plausible, other explanations are too.  As large as YouTube is, it is just a part of an even larger information ecosystem that includes the entire web, along with TV and radio. Thus, the growing engagement with radical content on YouTube may simply reflect a more general trend driven by a complicated combination of causes, both technological and sociological, that extend beyond the scope of the platform's algorithms and boundaries. 

In order to disambiguate between these explanations, we performed three additional analyses. First, we examined whether YouTube consumption is aberrant relative to off-platform consumption of similar content. Second, we analyzed the exact pathways by which users encountered political content on YouTube, thereby placing an upper bound on the fraction of views that could have been caused by the recommender. Finally, we checked whether political content is more likely to be consumed later in a user session, when the recommendation algorithm has had more opportunities to recommend content.

% here again we need to be general, when we enumerate the result we can say far right
Although none of these analyses on its own can rule out--or in--the causal effect of the recommendation engine, the strongest evidence for such an effect would be (a) higher on-platform consumption of radical political content than off-platform, (b) arrival at radical political content dominated by immediately previous video views (thereby implicating the recommender), and (c) increasing frequency of radical political content toward the end of a session, especially a long session. By contrast, the strongest evidence for outside influences would be (a) high correlation between on- and off-platform tastes, (b) arrival dominated by referral from outside websites or search, and (c) no increase in frequency over sessions, even long ones. 

%%%%%%%%%%%%%%%%%%%%
%% FIGURE 6
\begin{figure}[tb!]
\centering
\includegraphics[width=5cm]{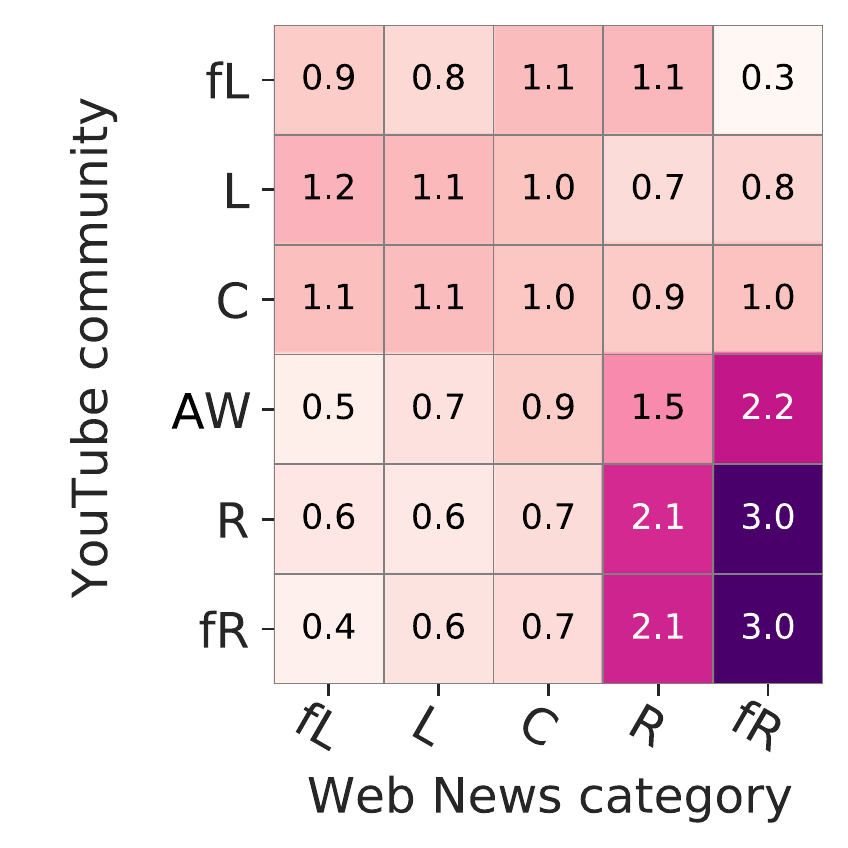}
\caption{Risk ratio of consumption from category $j$ on web ($j\in\{{\rm{fL, L, C, R, fR}}\}$) for users inside each community $i\in\{{\rm{fL, L, C, AW, R, fR}}\}$ on YouTube. Users of far right, right, and AW are more likely than random YouTube users to consume from right and far-right content on web.}
\label{fig:riskratio}
\end{figure}
%%%%%%%%%%%%%%%%%%%%

%%%%%%%%%%%%%%%%%%%%
%% TABLE 2
\begin{table*}[tb!]%[tbhp]
\centering
\caption{Distribution of the entry points of videos within each category. Video URLs can start from a YouTube homepage, a search (on/off platform), a YouTube user/channel, another YouTube video, or an external URL. YouTube recommendation (video entry point) have a bigger role for left (40\%) than any other group.}
\begin{tabular}{ccccccc}
\hline \hline
category & YouTube homepage & search & YouTube user/ channel & YouTube video & external URLs & other \\ \hline \hline
fL & $9.18$  & $9.7$  & $3.5$  & $27.71$  & $47.77$  & $2.15$  \\ \hline
L & $10.72$  & $10.45$  & $2.72$  & $40.06$  & $33.72$  & $2.33$  \\ \hline
C & $9.36$  & $13.76$  & $1.66$  & $31.96$  & $40.17$  & $3.08$  \\ \hline
AW & $11.98$  & $7.46$  & $3.63$  & $38.62$  & $35.51$  & $2.8$  \\ \hline
R & $7.19$  & $9.12$  & $4.23$  & $37.67$  & $40.19$  & $1.6$  \\ \hline
fR & $7.85$  & $6.36$  & $6.85$  & $35.8$  & $41.29$  & $1.86$  \\ \hline
\bottomrule
\end{tabular}\label{table:entrance}
\end{table*}
%%%%%%%%%%%%%%%%%%%%

\emph{On- vs. Off-platform.} 
To check for differences in on- vs. off-platform consumption, we compared the YouTube consumption of members of our six previously identified communities with their consumption of non-YouTube websites classified according to the far-left to far-right web categories. To label websites, we first identified news domains using Nielsen's classification scheme, which distinguishes between themes such as entertainment, travel, finance, etc. Out of all web domains accessed by individuals in our data set, $3,362$ were in the news category. We then used the partisan audience bias score provided by Ref.~\cite{robertson2018auditing} to bucket the news domains into the five political labels---fL,~L,~C,~R,~fR---that we used for YouTube channels above.% (see SI Appendix, section C for more details). 
 ~Although some anti-woke YouTube channel owners may also host external websites, there is no equivalent of the anti-woke category in Ref.~\cite{robertson2018auditing}; thus, we do not include it as a category for external websites.
 
To examine how the consumption pattern of users inside YouTube communities is associated with their web content consumption, Fig.~\ref{fig:riskratio} shows the risk ratio $RR_{(i,j)}=\frac{P_{ij}}{P_j}$ for each YouTube community $i\in\{{\rm{fL, L, C, AW, R, fR}}\}$, and each web category $j\in\{{\rm{fL, L, C, R, fR}}\}$, where $P_{ij}$ is probability of consumption from category $j$ on web, given users belong to community $i$ on YouTube, and $P_{j}$ is probability of consuming from category $j$ on web for random YouTube users. 
Fig.~\ref{fig:riskratio} shows two main results. First, members of the right and far-right communities are more than twice as likely to consume right-leaning content and three times as likely to consume far-right content outside of YouTube compared with an average user. Second, external consumption by members of the anti-woke community is biased in a very similar way to that of right and far-right members (1.5$\times$ for right-leaning content and 2.2$\times$ for far-right). In other words, even if anti-woke channel owners do not see themselves as associating with right or far-right ideologies, their viewers do. 
Finally, we note that while these results do not rule out that recommendations are driving engagement for the heaviest consumers, they are strongly consistent with the explanation that consumption of radical content on YouTube \duncan{is at least in part a reflection of user preferences}
%is highly correlated with its consumption across the web in general
~\cite{munger2020right,wilson2020cross}.% (see also SI Appendix Fig.~S10 for more details on correlations between internal and external consumption). 

\emph{Referral mode.}
Next, we explore how users encounter YouTube content by identifying ``referral'' pages, which we define as the page visited immediately prior to each YouTube video. We then classify referral pages as belonging to one of six categories (1) the YouTube homepage, (2) search (inside YouTube or external search engines), (3) a YouTube user/channel, (4) another YouTube video, (5) an external (non-YouTube) URL, and (6) other miscellaneous YouTube pages, such as feed, history, etc.  
 Table \ref{table:entrance} shows that while 36\% of far-right videos are preceded by another video,
  nearly 55\% of referrals come from one of: external URLs (41\%); the YouTube homepage (8\%); search queries (6\%).% (see SI appendix as a robustness check for more strict definitions of news consumers, Fig. S15). 
  ~Moreover, focusing on the subset of videos that are watched immediately after a user visits a news web domain, we find that approximately $50\%$ of far-right/right videos and more than $30\%$ of anti-woke videos are begun after visiting a right or far-right news domain such as ``foxnews.com,'' ``breitbart.com,'' and ``infowars.com,''. In contrast, if the video is from a far-left, left or center channel, it is highly likely ($70\%$) that the external entrance domain belongs to the center news bucket, which indicates domains like ``nytimes.com'' and ``bloomberg.com.''% (SI Appendix, Figs. S11-S12).
% politico
%%%%%%%%%%%%%%%%%%%%
%% FIGURE 7
\begin{figure*}[tb!]
\centering
\includegraphics[width=17cm]{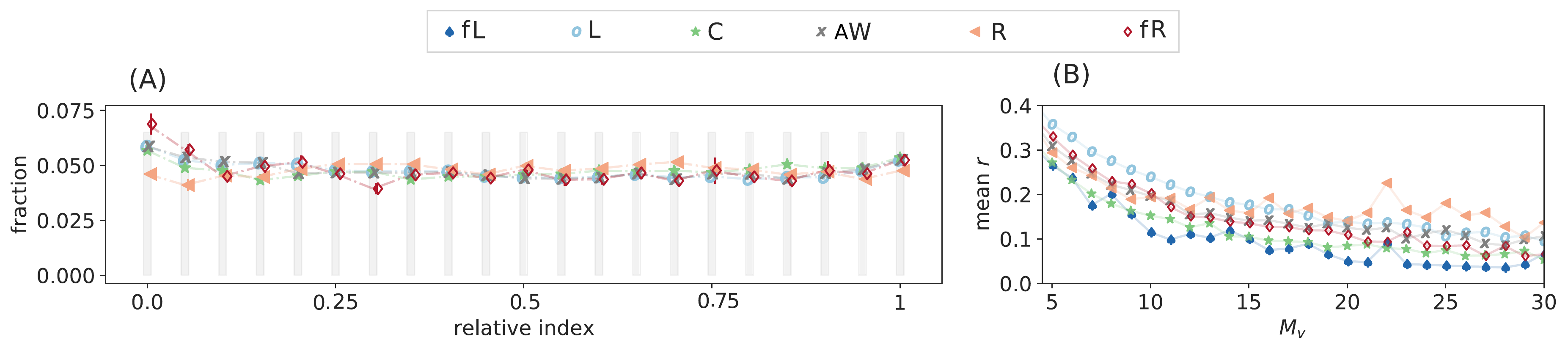}
\caption{ (A) Mean and standard deviation of fractions of videos as a function of normalized relative indices across session definitions for each category $k$, $k\in\{{\rm{L,~C,~AW,~R,~fR}}\}$, for sessions with length $M_k \geq 20$. (B) Average $r$, where $r$ is the fraction of videos of category $k$, $k\in\{{\rm{fL,~L,~C,~AW,~R,~fR}}\}$, for session definition $\delta=10$ and $\gamma =60$ minutes. Sessions with length $M_v\geq 30$ (2\% of the sessions) are dropped for better visualization.}
\label{fig:session_index_longer}
\end{figure*}
%%%%%%%%%%%%%%%%%%%%

\paragraph{Session analysis.} 
Although our data do not reveal which videos are being recommended to a user, if the recommendation algorithm is systematically promoting \duncan{a certain type of content, we would expect to observe increased viewership of the corresponding category (a) over the course of a session; and (b) as session length increases.} 
For example, if a user who initiates a session by viewing centrist or right-leaning videos is systematically directed toward far-right content, we would expect to observe a relatively higher frequency of far-right videos towards the end of the session. Moreover, because algorithmic recommendations have more opportunities to influence viewing choices as session length increases, we would expect to see higher relative frequency of far-right videos in longer sessions than shorter ones. 
Conversely, if we observe no increase in the relative frequency of far-right videos either over the course of a session or with session length, it would be evidence inconsistent with the \duncan{claim} that the recommender is driving users toward radical content. 

To test these hypotheses, we assigned each video with political label $k\in\{{\rm{fL,~L,~C,~AW,~R,~fR}}\}$ an index $i^{(k)} \in \{1,..,M_v\}$ where $M_v$ is the number of videos in the sessions. We then normalized the indices  $i^{(k)}_{{\rm{norm}}}=\frac{i^{(k)}-1}{M_v-1}$, such that $i^{(k)}_{{\rm{norm}}} \in [0,1]$, meaning that zero indicates the first video and one indicates the final video of a session. 
Fig.~\ref{fig:session_index_longer}A shows mean and standard deviation of the fraction of videos with normalized index $i^{(k)}_{{\rm{norm}}}$ for sessions of length $M_v \geq 20 $, for each category $k$, $k\in\{{\rm{L,~C,~AW,~R,~fR}}\}$ across different session definitions.% (SI Appendix, Table S16). 
~The fraction of videos from the far-left category is too small to provide clear statistical results (Fig.~\ref{fig:session_index_longer}B), and hence we dropped it from Fig.~\ref{fig:session_index_longer}A.
In all remaining cases, we find a nearly uniform distribution with an entropy deviating only slightly from that of a perfectly uniform distribution. %(SI Appendix, Table S17%\ref{table:session_relative_index})
~For longer sessions, there is a slightly higher density closer to the relative index 0 for far-right videos, precisely the opposite of what we would expect if the recommender were responsible.% (see SI Appendix, Figs. S19-S20 and Table S17 %\ref{fig:session_index1}, \ref{fig:session_index2} and \ref{fig:session_index_dist_S}, and Table \ref{table:session_relative_index} 
%for more details and robustness checks). 
~Complementing the within-session analysis, Fig.~\ref{fig:session_index_longer}B shows the average frequency of content categories as a function of session length. All six content categories show overall decreasing frequency, suggesting that longer sessions are increasingly devoted to non-news content. 
%One possible exception to this general pattern is an uptick in frequency of right videos at very long (>20 videos) session lengths; however, the data is sufficiently sparse that the uptick could simply be noise. 
\duncan{More specifically, we see no evidence that far-right content is more likely to be consumed in longer sessions---in fact, we observe precisely the opposite.}

\section*{Discussion}
The internet has fundamentally altered the production and consumption of political news content. On the production side, it has dramatically reduced the barriers to entry for would-be publishers of news, leading to a proliferation of small and often unreliable sources of information~\cite{AlexStamos}. On the consumption side, search and recommendation engines make even marginal actors easily discoverable, allowing them to build large, highly engaged audiences at a low cost. And, the sheer size of online platforms---Facebook and YouTube each have more than $2$ billion active users per month---tends to scale up the effect of any flaws in their algorithmic design or content moderation policies.

%In this paper we have studied two possible concerns about the consumption of far-right and anti-woke content that previous work has raised with respect to YouTube.
%\dmr{together they account for roughly one third of total online news watch time} (SI Appendix, Fig. S8%\ref{fig:Median_YouTube_news_consumption}
Here, we evaluated first whether YouTube, via its recommendation algorithm, drives attention to far-right content such as white supremacist ideologies and QAnon conspiracies, and hence is effectively radicalizing its users~\cite{ribeiro2019auditing,munger2020right,ledwich2019algorithmic,faddoul2020longitudinal,alfano2020technologically,cho2020search}. %Second, that YouTube hosts a large and thriving community of ``anti woke'' channels that, while ostensibly non-ideological, effectively act as gateways to far-right content~\cite{lewis2018alternative}.  
Additionally, we investigate whether and how the large and growing community of ``anti woke'' channels act as a gateway to far-right content~\cite{lewis2018alternative}.  
%The potential for this audience to be exposed to extreme and conspiratorial content is substantial, especially if the platform itself is responsible, via its recommendation algorithms~\cite{ribeiro2019auditing,munger2020right,ledwich2019algorithmic,faddoul2020longitudinal,alfano2020technologically,cho2020search}.   
We have investigated these possibilities by analyzing the detailed news consumption of more than 300,000 YouTube users who watched more than 20 million videos, with nearly half a million videos spanning the political spectrum over a four year time period. 
%Our results show the existence communities with distinct political news habits on YouTube, in which users predominantly consume videos from one political category (Fig. \ref{fig:archetypes}). 
%This result stands in contrast with generic web news consumption, for which previous studies have generally failed to find such \hh{echo chambers --> communities}~\cite{flaxman2016filter,zuiderveen2016should,guess2018almost}, suggesting that YouTube may play a special role in the consumption of digital news.

Our results show that a community of users who predominantly consume content produced by far-right channels does exist, \duncan{and while larger than the corresponding far-left community, it is small compared with centrist, left-leaning or right-leaning communities and not increasing in size over the time period of our study. 
%We also find little evidence that consumption of far-right content is driven by YouTube's recommendation system. 
%Overall, in fact, we find that centrist and left-leaning content dominates news consumption on YouTube and 
Moreover, we find that on-platform consumption of far-right content correlates highly with off-platform consumption of similar content; that users are roughly twice as likely to arrive at a far-right video from some source other than a previous YouTube video (e.g. search, an external website, the home page); and that far-right videos are no more likely to be viewed toward the end of sessions or in longer sessions. While none of this evidence can rule out the recommendation system as a cause of traffic to far-right content, it is more consistent with users simply having a preference for the content they consume.} 
%Among these communities, the left-leaning one is easily the largest both in terms of size and watch time (Fig.~\ref{fig:trends}). Moreover, the overall trend is towards an increasingly left-leaning and centrist distribution of users (Fig.~\ref{fig:clustering}). 

\duncan{We also find that the anti-woke community, while still small compared with left and centrist communities, is larger than the far-right and is growing over time, both in size and engagement. %As with far-right content, we find no evidence that anti-woke content benefits preferentially from the recommendation system. We do find evidence that the anti-woke community draws members from the far-right more than from any other political community, and that anti-woke members show an affinity for far-right content off-platform. 
We find evidence that the anti-woke community draws members from the far-right more than from any other political community, and that anti-woke members show an affinity for far-right content off-platform. On the other hand, when they leave, anti-woke members are more likely to move to left, center, and right than far right. Thus, while there do seem to be links between the anti-woke and far-right communities in terms of the content they consume, the hypothesized role of anti-woke channels as a gateway to far right is not supported. Rather it seems more accurate to describe anti woke as an increasingly popular---and sticky---category of its own. %Whether or not that should be regarded as problematic is beyond the scope of this paper.
 The implications of this fact are beyond the scope of this paper and left for future work to explore.}

Overall, our findings suggest that YouTube---while clearly an important destination for producers and consumers of
political content---is best understood as part of a larger ecosystem for distributing, discovering, and consuming political content~\cite{munger2020right,wilson2020cross}. Although much about the dynamics of this large ecosystem remains to be understood, it is plausible to think of YouTube as one of many ``libraries'' of content---albeit an especially large and prominent one---to which search engines and other websites (e.g., Rush Limbaugh’s blog or breitbart.com) can direct their users~\cite{wilson2020cross}. Once they have arrived at the ``library,'' users may continue to browse other similar content, and YouTube presumably exerts some control over these subsequent choices via its recommendations. 
Notably, $80\%$ of sessions are length one and that $55\%$ of videos are in sessions of no more than four videos. Both the majority of sessions and overall consumption, in other words, reflect the tastes and intentions that users entered with, and that they also exhibit in their general web browsing behavior.
To the extent that the growing consumption of radical political content is a social problem, our findings suggest that it is a much broader phenomenon than simply the policies and algorithmic properties of a single platform, even one as large as YouTube.

Our analysis comes with important limitations. First, our method of classifying content in terms of channel categories is an imperfect proxy for the content of individual videos. Just because a particular channel produces a substantial amount of far-right content, and hence could be legitimately classified as ``far right'' in our taxonomy, does not mean that every video promotes far-right ideology or is even political in nature. Future work could adopt a ``content-based'' classification system that could identify radical content more precisely. Second, 
while our panel-based method has the advantage of measuring consumption directly, it does not allow us to see videos that were recommended but not chosen. Fully reconstructing the decision processes of users would therefore require a combination of panel and platform data. Third, our data only includes desktop browsing, and hence reflects the behaviors of people who tend to use desktops for web browsing. Using recently acquired mobile panel that tracks total time spent on YouTube but not detailed in-app usage, we are able to compare the fraction of mobile/desktop panel users who access YouTube at least once per month, and also the median consumption time per mobile/desktop user. Roughly two times as many mobile users as desktop users visit YouTube at least once a month (40\% vs 20\%). % but desktop users spend somewhat more time on the site (right panel). 
Although we cannot rule out that consumption of radical political content is higher on mobile vs. desktop, other recent analysis of mobile device usage found that online news consumption is greater via laptop/desktop browsers \cite{allen2020evaluating}. %, \amir{suggesting that it is not}. Moreover, these results also strongly suggest that mobile session are shorter, and hence that the effect of recommendations will be even weaker than we detect. 
~Better integration of desktop with mobile consumption presents a major challenge for future work. Fourth, while our sample of videos is large and encompasses most popular channels, we cannot guarantee that all content of interest has been included. Future work would therefore benefit from yet larger and more comprehensive samples, both of videos and channels. Fifth, because our sample is retrospective, roughly 20\% of videos had been unavailable when we attempted to access their labels using the YouTube API. Although we were able to impute these labels using a classifier trained on a small sample of videos for which labels were available from other sources, it would be preferable in future work to obtain the true labels in closer to real time. Sixth, our method for identifying referral pages does not account for the possibility that users move between multiple ``tabs'' on their web browsers, all open simultaneously. Moreover, viewing credit in the Nielsen panel is only assigned to videos that are playing in the foreground, allowing for the possibility that other videos are playing automatically on background tabs. As a result, some videos that we have attributed to external websites may have in fact been suggested by the recommender in a background tab. The results presented in Table \ref{table:entrance} thus should be viewed as an upper bound for ``external'' entrances and lower bound for ``video'' entrances.  Finally, we emphasize that our analysis is intended to address systematic effects and hence applies only at the population level. It  says little about the role of specific YouTube content and the possible radicalization of individuals or small groups of individuals.% (\hl{SI Appendix, section H <- David kept it in response letter, pending on Duncan's approval}). % , but it is not YouTube that is doing the radicalizing.}

In addition to addressing these limitations, we hope that future work will address the broader issue of shifting consumption patterns that are driven by ``cord cutting'' and other technology-dependent changes in consumer behaviour. Although recent work has shown that television remains by far the dominant source of news for most Americans~\cite{allen2020evaluating}, our results suggest that online video content---on YouTube in particular---is increasingly competing with cable and network news for viewers. If so, and if the ``market'' for online video news is one in which small, low-quality purveyors of hyperpartisan, conspiratorial, or otherwise misleading content can compete with established brands, the combination of high engagement and large audience size 
may both fragment and complicate efforts to understand political content consumption and its social impact. While misinformation research has to date focused on text-heavy platforms such as Twitter and Facebook, we suggest that video deserves equal attention.  

\vspace{-2mm}
\begin{acknowledgements}
\textbf{Acknowledgments}: We are grateful to Harmony Laboratories for engineering and financial support and to Nielsen for access to panel data. We are also grateful to Kevin Munger, Manoel Horta Ribeiro, and Mark Ledwich for sharing their datasets with us and Daniel Muise, Keith Golden, Yue Chen, and Tushar Chandra for help with data preparation. Additional financial support for this research was provided by the Nathan Cummings Foundation (Grants 17-07331 and 18-08129) and the Carnegie Corporation of New York (Grant G-F-20-57741).
\end{acknowledgements}

% Bibliography

%\bibliography{references}

\begin{thebibliography}{40}%
\makeatletter
\providecommand \@ifxundefined [1]{%
 \@ifx{#1\undefined}
}%
\providecommand \@ifnum [1]{%
 \ifnum #1\expandafter \@firstoftwo
 \else \expandafter \@secondoftwo
 \fi
}%
\providecommand \@ifx [1]{%
 \ifx #1\expandafter \@firstoftwo
 \else \expandafter \@secondoftwo
 \fi
}%
\providecommand \natexlab [1]{#1}%
\providecommand \enquote  [1]{``#1''}%
\providecommand \bibnamefont  [1]{#1}%
\providecommand \bibfnamefont [1]{#1}%
\providecommand \citenamefont [1]{#1}%
\providecommand \href@noop [0]{\@secondoftwo}%
\providecommand \href [0]{\begingroup \@sanitize@url \@href}%
\providecommand \@href[1]{\@@startlink{#1}\@@href}%
\providecommand \@@href[1]{\endgroup#1\@@endlink}%
\providecommand \@sanitize@url [0]{\catcode `\\12\catcode `\$12\catcode
  `\&12\catcode `\#12\catcode `\^12\catcode `\_12\catcode `\%12\relax}%
\providecommand \@@startlink[1]{}%
\providecommand \@@endlink[0]{}%
\providecommand \url  [0]{\begingroup\@sanitize@url \@url }%
\providecommand \@url [1]{\endgroup\@href {#1}{\urlprefix }}%
\providecommand \urlprefix  [0]{URL }%
\providecommand \Eprint [0]{\href }%
\providecommand \doibase [0]{http://dx.doi.org/}%
\providecommand \selectlanguage [0]{\@gobble}%
\providecommand \bibinfo  [0]{\@secondoftwo}%
\providecommand \bibfield  [0]{\@secondoftwo}%
\providecommand \translation [1]{[#1]}%
\providecommand \BibitemOpen [0]{}%
\providecommand \bibitemStop [0]{}%
\providecommand \bibitemNoStop [0]{.\EOS\space}%
\providecommand \EOS [0]{\spacefactor3000\relax}%
\providecommand \BibitemShut  [1]{\csname bibitem#1\endcsname}%
\let\auto@bib@innerbib\@empty
%</preamble>
\bibitem [{\citenamefont {Iyengar}\ \emph {et~al.}(2019)\citenamefont
  {Iyengar}, \citenamefont {Lelkes}, \citenamefont {Levendusky}, \citenamefont
  {Malhotra},\ and\ \citenamefont {Westwood}}]{iyengar2019origins}%
  \BibitemOpen
  \bibfield  {author} {\bibinfo {author} {\bibfnamefont {S.}~\bibnamefont
  {Iyengar}}, \bibinfo {author} {\bibfnamefont {Y.}~\bibnamefont {Lelkes}},
  \bibinfo {author} {\bibfnamefont {M.}~\bibnamefont {Levendusky}}, \bibinfo
  {author} {\bibfnamefont {N.}~\bibnamefont {Malhotra}}, \ and\ \bibinfo
  {author} {\bibfnamefont {S.~J.}\ \bibnamefont {Westwood}},\ }\href@noop {}
  {\bibfield  {journal} {\bibinfo  {journal} {Annual Review of Political
  Science}\ }\textbf {\bibinfo {volume} {22}},\ \bibinfo {pages} {129}
  (\bibinfo {year} {2019})}\BibitemShut {NoStop}%
\bibitem [{\citenamefont {Jones}(2015)}]{jones2015declining}%
  \BibitemOpen
  \bibfield  {author} {\bibinfo {author} {\bibfnamefont {D.~R.}\ \bibnamefont
  {Jones}},\ }in\ \href@noop {} {\emph {\bibinfo {booktitle} {The Forum}}},\
  Vol.~\bibinfo {volume} {13}\ (\bibinfo {organization} {De Gruyter},\ \bibinfo
  {year} {2015})\ pp.\ \bibinfo {pages} {375--394}\BibitemShut {NoStop}%
\bibitem [{\citenamefont {Tucker}\ \emph {et~al.}(2018)\citenamefont {Tucker},
  \citenamefont {Guess}, \citenamefont {Barber{\'a}}, \citenamefont {Vaccari},
  \citenamefont {Siegel}, \citenamefont {Sanovich}, \citenamefont {Stukal},\
  and\ \citenamefont {Nyhan}}]{tucker2018social}%
  \BibitemOpen
  \bibfield  {author} {\bibinfo {author} {\bibfnamefont {J.~A.}\ \bibnamefont
  {Tucker}}, \bibinfo {author} {\bibfnamefont {A.}~\bibnamefont {Guess}},
  \bibinfo {author} {\bibfnamefont {P.}~\bibnamefont {Barber{\'a}}}, \bibinfo
  {author} {\bibfnamefont {C.}~\bibnamefont {Vaccari}}, \bibinfo {author}
  {\bibfnamefont {A.}~\bibnamefont {Siegel}}, \bibinfo {author} {\bibfnamefont
  {S.}~\bibnamefont {Sanovich}}, \bibinfo {author} {\bibfnamefont
  {D.}~\bibnamefont {Stukal}}, \ and\ \bibinfo {author} {\bibfnamefont
  {B.}~\bibnamefont {Nyhan}},\ }\href@noop {} {\bibfield  {journal} {\bibinfo
  {journal} {Political polarization, and political disinformation: a review of
  the scientific literature (March 19, 2018)}\ } (\bibinfo {year}
  {2018})}\BibitemShut {NoStop}%
\bibitem [{\citenamefont {Conover}\ \emph {et~al.}(2011)\citenamefont
  {Conover}, \citenamefont {Ratkiewicz}, \citenamefont {Francisco},
  \citenamefont {Gon{\c{c}}alves}, \citenamefont {Menczer},\ and\ \citenamefont
  {Flammini}}]{conover2011political}%
  \BibitemOpen
  \bibfield  {author} {\bibinfo {author} {\bibfnamefont {M.~D.}\ \bibnamefont
  {Conover}}, \bibinfo {author} {\bibfnamefont {J.}~\bibnamefont {Ratkiewicz}},
  \bibinfo {author} {\bibfnamefont {M.~R.}\ \bibnamefont {Francisco}}, \bibinfo
  {author} {\bibfnamefont {B.}~\bibnamefont {Gon{\c{c}}alves}}, \bibinfo
  {author} {\bibfnamefont {F.}~\bibnamefont {Menczer}}, \ and\ \bibinfo
  {author} {\bibfnamefont {A.}~\bibnamefont {Flammini}},\ }\href@noop {}
  {\bibfield  {journal} {\bibinfo  {journal} {The International AAAI Conference
  on Web and Social Media ({ICWSM})}\ }\textbf {\bibinfo {volume} {133}},\
  \bibinfo {pages} {89} (\bibinfo {year} {2011})}\BibitemShut {NoStop}%
\bibitem [{\citenamefont {Del~Vicario}\ \emph {et~al.}(2016)\citenamefont
  {Del~Vicario}, \citenamefont {Vivaldo}, \citenamefont {Bessi}, \citenamefont
  {Zollo}, \citenamefont {Scala}, \citenamefont {Caldarelli},\ and\
  \citenamefont {Quattrociocchi}}]{del2016echo}%
  \BibitemOpen
  \bibfield  {author} {\bibinfo {author} {\bibfnamefont {M.}~\bibnamefont
  {Del~Vicario}}, \bibinfo {author} {\bibfnamefont {G.}~\bibnamefont
  {Vivaldo}}, \bibinfo {author} {\bibfnamefont {A.}~\bibnamefont {Bessi}},
  \bibinfo {author} {\bibfnamefont {F.}~\bibnamefont {Zollo}}, \bibinfo
  {author} {\bibfnamefont {A.}~\bibnamefont {Scala}}, \bibinfo {author}
  {\bibfnamefont {G.}~\bibnamefont {Caldarelli}}, \ and\ \bibinfo {author}
  {\bibfnamefont {W.}~\bibnamefont {Quattrociocchi}},\ }\href@noop {}
  {\bibfield  {journal} {\bibinfo  {journal} {Scientific Reports}\ }\textbf
  {\bibinfo {volume} {6}},\ \bibinfo {pages} {37825} (\bibinfo {year}
  {2016})}\BibitemShut {NoStop}%
\bibitem [{\citenamefont {Grover}\ \emph {et~al.}(2019)\citenamefont {Grover},
  \citenamefont {Kar}, \citenamefont {Dwivedi},\ and\ \citenamefont
  {Janssen}}]{grover2019polarization}%
  \BibitemOpen
  \bibfield  {author} {\bibinfo {author} {\bibfnamefont {P.}~\bibnamefont
  {Grover}}, \bibinfo {author} {\bibfnamefont {A.~K.}\ \bibnamefont {Kar}},
  \bibinfo {author} {\bibfnamefont {Y.~K.}\ \bibnamefont {Dwivedi}}, \ and\
  \bibinfo {author} {\bibfnamefont {M.}~\bibnamefont {Janssen}},\ }\href@noop
  {} {\bibfield  {journal} {\bibinfo  {journal} {Technological Forecasting and
  Social Change}\ }\textbf {\bibinfo {volume} {145}},\ \bibinfo {pages} {438}
  (\bibinfo {year} {2019})}\BibitemShut {NoStop}%
\bibitem [{\citenamefont {Bossetta}(2018)}]{bossetta2018digital}%
  \BibitemOpen
  \bibfield  {author} {\bibinfo {author} {\bibfnamefont {M.}~\bibnamefont
  {Bossetta}},\ }\href@noop {} {\bibfield  {journal} {\bibinfo  {journal}
  {Journalism \& {M}ass {C}ommunication {Q}uarterly}\ }\textbf {\bibinfo
  {volume} {95}},\ \bibinfo {pages} {471} (\bibinfo {year} {2018})}\BibitemShut
  {NoStop}%
\bibitem [{\citenamefont {Alizadeh}\ \emph {et~al.}(2020)\citenamefont
  {Alizadeh}, \citenamefont {Shapiro}, \citenamefont {Buntain},\ and\
  \citenamefont {Tucker}}]{alizadeh2020content}%
  \BibitemOpen
  \bibfield  {author} {\bibinfo {author} {\bibfnamefont {M.}~\bibnamefont
  {Alizadeh}}, \bibinfo {author} {\bibfnamefont {J.~N.}\ \bibnamefont
  {Shapiro}}, \bibinfo {author} {\bibfnamefont {C.}~\bibnamefont {Buntain}}, \
  and\ \bibinfo {author} {\bibfnamefont {J.~A.}\ \bibnamefont {Tucker}},\
  }\href@noop {} {\bibfield  {journal} {\bibinfo  {journal} {Science Advances}\
  }\textbf {\bibinfo {volume} {6}},\ \bibinfo {pages} {eabb5824} (\bibinfo
  {year} {2020})}\BibitemShut {NoStop}%
\bibitem [{\citenamefont {Aral}\ and\ \citenamefont
  {Eckles}(2019)}]{aral2019protecting}%
  \BibitemOpen
  \bibfield  {author} {\bibinfo {author} {\bibfnamefont {S.}~\bibnamefont
  {Aral}}\ and\ \bibinfo {author} {\bibfnamefont {D.}~\bibnamefont {Eckles}},\
  }\href@noop {} {\bibfield  {journal} {\bibinfo  {journal} {Science}\ }\textbf
  {\bibinfo {volume} {365}},\ \bibinfo {pages} {858} (\bibinfo {year}
  {2019})}\BibitemShut {NoStop}%
\bibitem [{\citenamefont {Lazer}\ \emph {et~al.}(2018)\citenamefont {Lazer},
  \citenamefont {Baum}, \citenamefont {Benkler}, \citenamefont {Berinsky},
  \citenamefont {Greenhill}, \citenamefont {Menczer}, \citenamefont {Metzger},
  \citenamefont {Nyhan}, \citenamefont {Pennycook}, \citenamefont {Rothschild}
  \emph {et~al.}}]{lazer2018science}%
  \BibitemOpen
  \bibfield  {author} {\bibinfo {author} {\bibfnamefont {D.~M.}\ \bibnamefont
  {Lazer}}, \bibinfo {author} {\bibfnamefont {M.~A.}\ \bibnamefont {Baum}},
  \bibinfo {author} {\bibfnamefont {Y.}~\bibnamefont {Benkler}}, \bibinfo
  {author} {\bibfnamefont {A.~J.}\ \bibnamefont {Berinsky}}, \bibinfo {author}
  {\bibfnamefont {K.~M.}\ \bibnamefont {Greenhill}}, \bibinfo {author}
  {\bibfnamefont {F.}~\bibnamefont {Menczer}}, \bibinfo {author} {\bibfnamefont
  {M.~J.}\ \bibnamefont {Metzger}}, \bibinfo {author} {\bibfnamefont
  {B.}~\bibnamefont {Nyhan}}, \bibinfo {author} {\bibfnamefont
  {G.}~\bibnamefont {Pennycook}}, \bibinfo {author} {\bibfnamefont
  {D.}~\bibnamefont {Rothschild}},  \emph {et~al.},\ }\href@noop {} {\bibfield
  {journal} {\bibinfo  {journal} {Science}\ }\textbf {\bibinfo {volume}
  {359}},\ \bibinfo {pages} {1094} (\bibinfo {year} {2018})}\BibitemShut
  {NoStop}%
\bibitem [{\citenamefont {Pennycook}\ and\ \citenamefont
  {Rand}(2019)}]{pennycook2019fighting}%
  \BibitemOpen
  \bibfield  {author} {\bibinfo {author} {\bibfnamefont {G.}~\bibnamefont
  {Pennycook}}\ and\ \bibinfo {author} {\bibfnamefont {D.~G.}\ \bibnamefont
  {Rand}},\ }\href@noop {} {\bibfield  {journal} {\bibinfo  {journal}
  {Proceedings of the National Academy of Sciences, {USA}}\ }\textbf {\bibinfo
  {volume} {116}},\ \bibinfo {pages} {2521} (\bibinfo {year}
  {2019})}\BibitemShut {NoStop}%
\bibitem [{\citenamefont {Roozenbeek}\ and\ \citenamefont {Van
  Der~Linden}(2019)}]{roozenbeek2019fake}%
  \BibitemOpen
  \bibfield  {author} {\bibinfo {author} {\bibfnamefont {J.}~\bibnamefont
  {Roozenbeek}}\ and\ \bibinfo {author} {\bibfnamefont {S.}~\bibnamefont {Van
  Der~Linden}},\ }\href@noop {} {\bibfield  {journal} {\bibinfo  {journal}
  {Journal of Risk Research}\ }\textbf {\bibinfo {volume} {22}},\ \bibinfo
  {pages} {570} (\bibinfo {year} {2019})}\BibitemShut {NoStop}%
\bibitem [{\citenamefont {Konitzer}\ \emph {et~al.}(2020)\citenamefont
  {Konitzer}, \citenamefont {Allen}, \citenamefont {Eckman}, \citenamefont
  {Howland}, \citenamefont {Mobius}, \citenamefont {Rothschild},\ and\
  \citenamefont {Watts}}]{konitzer2020measuring}%
  \BibitemOpen
  \bibfield  {author} {\bibinfo {author} {\bibfnamefont {T.}~\bibnamefont
  {Konitzer}}, \bibinfo {author} {\bibfnamefont {J.}~\bibnamefont {Allen}},
  \bibinfo {author} {\bibfnamefont {S.}~\bibnamefont {Eckman}}, \bibinfo
  {author} {\bibfnamefont {B.}~\bibnamefont {Howland}}, \bibinfo {author}
  {\bibfnamefont {M.~M.}\ \bibnamefont {Mobius}}, \bibinfo {author}
  {\bibfnamefont {D.~M.}\ \bibnamefont {Rothschild}}, \ and\ \bibinfo {author}
  {\bibfnamefont {D.}~\bibnamefont {Watts}},\ }\href@noop {} {\bibfield
  {journal} {\bibinfo  {journal} {Available at SSRN 3548690}\ } (\bibinfo
  {year} {2020})}\BibitemShut {NoStop}%
\bibitem [{\citenamefont {{Audrey Schomer}}(2020)}]{emarketer_youtube}%
  \BibitemOpen
  \bibfield  {author} {\bibinfo {author} {\bibnamefont {{Audrey Schomer}}},\
  }\href@noop {} {\enquote {\bibinfo {title} {{US} {YouTube} advertising
  2020},}\ }\bibinfo {howpublished}
  {\url{https://www.emarketer.com/content/us-youtube-advertising-2020}}
  (\bibinfo {year} {2020})\BibitemShut {NoStop}%
\bibitem [{\citenamefont {{Mansoor Iqbal}}(2021)}]{businessofapps_twitter}%
  \BibitemOpen
  \bibfield  {author} {\bibinfo {author} {\bibnamefont {{Mansoor Iqbal}}},\
  }\href@noop {} {\enquote {\bibinfo {title} {{Twitter} revenue and usage
  statistics (2020)},}\ }\bibinfo {howpublished}
  {\url{https://www.businessofapps.com/data/twitter-statistics/}} (\bibinfo
  {year} {2021})\BibitemShut {NoStop}%
\bibitem [{\citenamefont {Clark}\ and\ \citenamefont
  {Zaitsev}(2020)}]{clark2020understanding}%
  \BibitemOpen
  \bibfield  {author} {\bibinfo {author} {\bibfnamefont {S.}~\bibnamefont
  {Clark}}\ and\ \bibinfo {author} {\bibfnamefont {A.}~\bibnamefont
  {Zaitsev}},\ }\href@noop {} {\bibfield  {journal} {\bibinfo  {journal} {arXiv
  preprint arXiv:2010.09892}\ } (\bibinfo {year} {2020})}\BibitemShut {NoStop}%
\bibitem [{\citenamefont {{Kevin Roose}}(2019)}]{rabbit_hole}%
  \BibitemOpen
  \bibfield  {author} {\bibinfo {author} {\bibnamefont {{Kevin Roose}}},\
  }\href@noop {} {\enquote {\bibinfo {title} {The making of a {YouTube}
  radical},}\ }\bibinfo {howpublished}
  {\url{https://www.nytimes.com/interactive/2019/06/08/technology/youtube-radical.html}}
  (\bibinfo {year} {2019})\BibitemShut {NoStop}%
\bibitem [{\citenamefont {{Zeynep Tufekci}}(2017)}]{youtube_radical}%
  \BibitemOpen
  \bibfield  {author} {\bibinfo {author} {\bibnamefont {{Zeynep Tufekci}}},\
  }\href@noop {} {\enquote {\bibinfo {title} {Youtube, the great
  radicalizer},}\ }\bibinfo {howpublished}
  {\url{https://www.nytimes.com/2018/03/10/opinion/sunday/youtube-politics-radical.html}}
  (\bibinfo {year} {2017})\BibitemShut {NoStop}%
\bibitem [{\citenamefont {{James Bridle}}(2017)}]{wrong_internet}%
  \BibitemOpen
  \bibfield  {author} {\bibinfo {author} {\bibnamefont {{James Bridle}}},\
  }\href@noop {} {\enquote {\bibinfo {title} {Something is wrong on the
  internet},}\ }\bibinfo {howpublished}
  {\url{https://medium.com/@jamesbridle/something-is-wrong-on-the-internet-c39c471271d2}}
  (\bibinfo {year} {2017})\BibitemShut {NoStop}%
\bibitem [{\citenamefont {{Tom Whyman}}(2017)}]{whyRight}%
  \BibitemOpen
  \bibfield  {author} {\bibinfo {author} {\bibnamefont {{Tom Whyman}}},\
  }\href@noop {} {\enquote {\bibinfo {title} {Why the right is dominating
  {YouTube}.}}\ }\bibinfo {howpublished}
  {\url{https://www.vice.com/en_us/article/3dy7vb/why-the-right-is-dominating-youtube}}
  (\bibinfo {year} {2017})\BibitemShut {NoStop}%
\bibitem [{\citenamefont {Lewis}(2018)}]{lewis2018alternative}%
  \BibitemOpen
  \bibfield  {author} {\bibinfo {author} {\bibfnamefont {R.}~\bibnamefont
  {Lewis}},\ }\href@noop {} {\bibfield  {journal} {\bibinfo  {journal} {Data \&
  Society Research Institute}\ } (\bibinfo {year} {2018})}\BibitemShut
  {NoStop}%
\bibitem [{\citenamefont {Lewis}(2020)}]{lewis2020news}%
  \BibitemOpen
  \bibfield  {author} {\bibinfo {author} {\bibfnamefont {R.}~\bibnamefont
  {Lewis}},\ }\href@noop {} {\bibfield  {journal} {\bibinfo  {journal}
  {Television \& New Media}\ }\textbf {\bibinfo {volume} {21}},\ \bibinfo
  {pages} {201} (\bibinfo {year} {2020})}\BibitemShut {NoStop}%
\bibitem [{\citenamefont {{Perry Bacon}}(2021)}]{bacon_anti_woke}%
  \BibitemOpen
  \bibfield  {author} {\bibinfo {author} {\bibnamefont {{Perry Bacon}}},\
  }\href@noop {} {\enquote {\bibinfo {title} {Why attacking ‘cancel
  culture’ and ‘woke’ people is becoming the gop’s new political
  strategy},}\ }\bibinfo {howpublished}
  {\url{https://fivethirtyeight.com/features/why-attacking-cancel-culture-and-woke-people-is-becoming-the-gops-new-political-strategy/}}
  (\bibinfo {year} {2021})\BibitemShut {NoStop}%
\bibitem [{\citenamefont {Ribeiro}\ \emph {et~al.}(2020)\citenamefont
  {Ribeiro}, \citenamefont {Ottoni}, \citenamefont {West}, \citenamefont
  {Almeida},\ and\ \citenamefont {Meira~Jr}}]{ribeiro2019auditing}%
  \BibitemOpen
  \bibfield  {author} {\bibinfo {author} {\bibfnamefont {M.~H.}\ \bibnamefont
  {Ribeiro}}, \bibinfo {author} {\bibfnamefont {R.}~\bibnamefont {Ottoni}},
  \bibinfo {author} {\bibfnamefont {R.}~\bibnamefont {West}}, \bibinfo {author}
  {\bibfnamefont {V.~A.}\ \bibnamefont {Almeida}}, \ and\ \bibinfo {author}
  {\bibfnamefont {W.}~\bibnamefont {Meira~Jr}},\ }in\ \href@noop {} {\emph
  {\bibinfo {booktitle} {{P}roceedings of the 2020 {C}onference on {F}airness,
  {A}ccountability, and {T}ransparency}}}\ (\bibinfo {year} {2020})\ pp.\
  \bibinfo {pages} {131--141}\BibitemShut {NoStop}%
\bibitem [{\citenamefont {{Maryam Mohsin}}(2020)}]{youtub_stats}%
  \BibitemOpen
  \bibfield  {author} {\bibinfo {author} {\bibnamefont {{Maryam Mohsin}}},\
  }\href@noop {} {\enquote {\bibinfo {title} {10 {YouTube} stats every marketer
  should know in 2020},}\ }\bibinfo {howpublished}
  {\url{https://www.oberlo.com/blog/youtube-statistics}} (\bibinfo {year}
  {2020})\BibitemShut {NoStop}%
\bibitem [{\citenamefont {Cho}\ \emph {et~al.}(2020)\citenamefont {Cho},
  \citenamefont {Ahmed}, \citenamefont {Hilbert}, \citenamefont {Liu},\ and\
  \citenamefont {Luu}}]{cho2020search}%
  \BibitemOpen
  \bibfield  {author} {\bibinfo {author} {\bibfnamefont {J.}~\bibnamefont
  {Cho}}, \bibinfo {author} {\bibfnamefont {S.}~\bibnamefont {Ahmed}}, \bibinfo
  {author} {\bibfnamefont {M.}~\bibnamefont {Hilbert}}, \bibinfo {author}
  {\bibfnamefont {B.}~\bibnamefont {Liu}}, \ and\ \bibinfo {author}
  {\bibfnamefont {J.}~\bibnamefont {Luu}},\ }\href@noop {} {\bibfield
  {journal} {\bibinfo  {journal} {Journal of Broadcasting \& Electronic Media}\
  ,\ \bibinfo {pages} {1}} (\bibinfo {year} {2020})}\BibitemShut {NoStop}%
\bibitem [{\citenamefont {Munger}\ and\ \citenamefont
  {Phillips}(2020)}]{munger2020right}%
  \BibitemOpen
  \bibfield  {author} {\bibinfo {author} {\bibfnamefont {K.}~\bibnamefont
  {Munger}}\ and\ \bibinfo {author} {\bibfnamefont {J.}~\bibnamefont
  {Phillips}},\ }\href@noop {} {\bibfield  {journal} {\bibinfo  {journal} {The
  International Journal of Press/Politics}\ ,\ \bibinfo {pages}
  {1940161220964767}} (\bibinfo {year} {2020})}\BibitemShut {NoStop}%
\bibitem [{\citenamefont {Ledwich}\ and\ \citenamefont
  {Zaitsev}(2019)}]{ledwich2019algorithmic}%
  \BibitemOpen
  \bibfield  {author} {\bibinfo {author} {\bibfnamefont {M.}~\bibnamefont
  {Ledwich}}\ and\ \bibinfo {author} {\bibfnamefont {A.}~\bibnamefont
  {Zaitsev}},\ }\href@noop {} {\bibfield  {journal} {\bibinfo  {journal} {First
  Monday, 25(3)}\ } (\bibinfo {year} {2019})}\BibitemShut {NoStop}%
\bibitem [{\citenamefont {Faddoul}\ \emph {et~al.}(2020)\citenamefont
  {Faddoul}, \citenamefont {Chaslot},\ and\ \citenamefont
  {Farid}}]{faddoul2020longitudinal}%
  \BibitemOpen
  \bibfield  {author} {\bibinfo {author} {\bibfnamefont {M.}~\bibnamefont
  {Faddoul}}, \bibinfo {author} {\bibfnamefont {G.}~\bibnamefont {Chaslot}}, \
  and\ \bibinfo {author} {\bibfnamefont {H.}~\bibnamefont {Farid}},\
  }\href@noop {} {\bibfield  {journal} {\bibinfo  {journal} {arXiv preprint
  arXiv:2003.03318}\ } (\bibinfo {year} {2020})}\BibitemShut {NoStop}%
\bibitem [{\citenamefont {Allen}\ \emph {et~al.}(2020)\citenamefont {Allen},
  \citenamefont {Howland}, \citenamefont {Mobius}, \citenamefont {Rothschild},\
  and\ \citenamefont {Watts}}]{allen2020evaluating}%
  \BibitemOpen
  \bibfield  {author} {\bibinfo {author} {\bibfnamefont {J.}~\bibnamefont
  {Allen}}, \bibinfo {author} {\bibfnamefont {B.}~\bibnamefont {Howland}},
  \bibinfo {author} {\bibfnamefont {M.}~\bibnamefont {Mobius}}, \bibinfo
  {author} {\bibfnamefont {D.}~\bibnamefont {Rothschild}}, \ and\ \bibinfo
  {author} {\bibfnamefont {D.~J.}\ \bibnamefont {Watts}},\ }\href@noop {}
  {\bibfield  {journal} {\bibinfo  {journal} {Science Advances}\ }\textbf
  {\bibinfo {volume} {6}},\ \bibinfo {pages} {eaay3539} (\bibinfo {year}
  {2020})}\BibitemShut {NoStop}%
\bibitem [{\citenamefont {Wilson}\ and\ \citenamefont
  {Starbird}(2020)}]{wilson2020cross}%
  \BibitemOpen
  \bibfield  {author} {\bibinfo {author} {\bibfnamefont {T.}~\bibnamefont
  {Wilson}}\ and\ \bibinfo {author} {\bibfnamefont {K.}~\bibnamefont
  {Starbird}},\ }\href@noop {} {\bibfield  {journal} {\bibinfo  {journal}
  {Harvard Kennedy School Misinformation Review}\ }\textbf {\bibinfo {volume}
  {1}} (\bibinfo {year} {2020})}\BibitemShut {NoStop}%
\bibitem [{\citenamefont {Lazer}(2020)}]{lazer2020studying}%
  \BibitemOpen
  \bibfield  {author} {\bibinfo {author} {\bibfnamefont {D.}~\bibnamefont
  {Lazer}},\ }\href@noop {} {\bibfield  {journal} {\bibinfo  {journal}
  {Proceedings of the National Academy of Sciences, USA}\ }\textbf {\bibinfo
  {volume} {117}},\ \bibinfo {pages} {21} (\bibinfo {year} {2020})}\BibitemShut
  {NoStop}%
\bibitem [{\citenamefont {Wu}\ \emph {et~al.}(2018)\citenamefont {Wu},
  \citenamefont {Rizoiu},\ and\ \citenamefont {Xie}}]{wu2018beyond}%
  \BibitemOpen
  \bibfield  {author} {\bibinfo {author} {\bibfnamefont {S.}~\bibnamefont
  {Wu}}, \bibinfo {author} {\bibfnamefont {M.-A.}\ \bibnamefont {Rizoiu}}, \
  and\ \bibinfo {author} {\bibfnamefont {L.}~\bibnamefont {Xie}},\ }in\
  \href@noop {} {\emph {\bibinfo {booktitle} {Twelfth International AAAI
  Conference on Web and Social Media}}}\ (\bibinfo {year} {2018})\BibitemShut
  {NoStop}%
\bibitem [{\citenamefont {{Sara Salinas}}(2018)}]{terminated_infowars}%
  \BibitemOpen
  \bibfield  {author} {\bibinfo {author} {\bibnamefont {{Sara Salinas}}},\
  }\href@noop {} {\enquote {\bibinfo {title} {{YouTube} removes {A}lex
  {J}ones’ page, following bans from {A}pple and {F}acebook},}\ }\bibinfo
  {howpublished}
  {\url{https://www.cnbc.com/2018/08/06/youtube-removes-alex-jones-account-following-earlier-bans.html/}}
  (\bibinfo {year} {2018})\BibitemShut {NoStop}%
\bibitem [{\citenamefont {Kumar}\ and\ \citenamefont
  {Tomkins}(2010)}]{kumar2010characterization}%
  \BibitemOpen
  \bibfield  {author} {\bibinfo {author} {\bibfnamefont {R.}~\bibnamefont
  {Kumar}}\ and\ \bibinfo {author} {\bibfnamefont {A.}~\bibnamefont
  {Tomkins}},\ }in\ \href@noop {} {\emph {\bibinfo {booktitle} {Proceedings of
  the 19th {I}nternational {C}onference on {W}orld {W}ide {W}eb}}}\ (\bibinfo
  {year} {2010})\ pp.\ \bibinfo {pages} {561--570}\BibitemShut {NoStop}%
\bibitem [{\citenamefont {Charrad}\ \emph {et~al.}(2014)\citenamefont
  {Charrad}, \citenamefont {Ghazzali}, \citenamefont {Boiteau},\ and\
  \citenamefont {Niknafs}}]{NbClust}%
  \BibitemOpen
  \bibfield  {author} {\bibinfo {author} {\bibfnamefont {M.}~\bibnamefont
  {Charrad}}, \bibinfo {author} {\bibfnamefont {N.}~\bibnamefont {Ghazzali}},
  \bibinfo {author} {\bibfnamefont {V.}~\bibnamefont {Boiteau}}, \ and\
  \bibinfo {author} {\bibfnamefont {A.}~\bibnamefont {Niknafs}},\ }\href
  {http://www.jstatsoft.org/v61/i06/} {\bibfield  {journal} {\bibinfo
  {journal} {Journal of Statistical Software}\ }\textbf {\bibinfo {volume}
  {61}},\ \bibinfo {pages} {1} (\bibinfo {year} {2014})}\BibitemShut {NoStop}%
\bibitem [{\citenamefont {Bail}\ \emph {et~al.}(2018)\citenamefont {Bail},
  \citenamefont {Argyle}, \citenamefont {Brown}, \citenamefont {Bumpus},
  \citenamefont {Chen}, \citenamefont {Hunzaker}, \citenamefont {Lee},
  \citenamefont {Mann}, \citenamefont {Merhout},\ and\ \citenamefont
  {Volfovsky}}]{Bail9216}%
  \BibitemOpen
  \bibfield  {author} {\bibinfo {author} {\bibfnamefont {C.~A.}\ \bibnamefont
  {Bail}}, \bibinfo {author} {\bibfnamefont {L.~P.}\ \bibnamefont {Argyle}},
  \bibinfo {author} {\bibfnamefont {T.~W.}\ \bibnamefont {Brown}}, \bibinfo
  {author} {\bibfnamefont {J.~P.}\ \bibnamefont {Bumpus}}, \bibinfo {author}
  {\bibfnamefont {H.}~\bibnamefont {Chen}}, \bibinfo {author} {\bibfnamefont
  {M.~B.~F.}\ \bibnamefont {Hunzaker}}, \bibinfo {author} {\bibfnamefont
  {J.}~\bibnamefont {Lee}}, \bibinfo {author} {\bibfnamefont {M.}~\bibnamefont
  {Mann}}, \bibinfo {author} {\bibfnamefont {F.}~\bibnamefont {Merhout}}, \
  and\ \bibinfo {author} {\bibfnamefont {A.}~\bibnamefont {Volfovsky}},\ }\href
  {\doibase 10.1073/pnas.1804840115} {\bibfield  {journal} {\bibinfo  {journal}
  {Proceedings of the National Academy of Sciences, USA}\ }\textbf {\bibinfo
  {volume} {115}},\ \bibinfo {pages} {9216} (\bibinfo {year} {2018})},\ \Eprint
  {http://arxiv.org/abs/https://www.pnas.org/content/115/37/9216.full.pdf}
  {https://www.pnas.org/content/115/37/9216.full.pdf} \BibitemShut {NoStop}%
\bibitem [{\citenamefont {Robertson}\ \emph {et~al.}(2018)\citenamefont
  {Robertson}, \citenamefont {Jiang}, \citenamefont {Joseph}, \citenamefont
  {Friedland}, \citenamefont {Lazer},\ and\ \citenamefont
  {Wilson}}]{robertson2018auditing}%
  \BibitemOpen
  \bibfield  {author} {\bibinfo {author} {\bibfnamefont {R.~E.}\ \bibnamefont
  {Robertson}}, \bibinfo {author} {\bibfnamefont {S.}~\bibnamefont {Jiang}},
  \bibinfo {author} {\bibfnamefont {K.}~\bibnamefont {Joseph}}, \bibinfo
  {author} {\bibfnamefont {L.}~\bibnamefont {Friedland}}, \bibinfo {author}
  {\bibfnamefont {D.}~\bibnamefont {Lazer}}, \ and\ \bibinfo {author}
  {\bibfnamefont {C.}~\bibnamefont {Wilson}},\ }\href@noop {} {\bibfield
  {journal} {\bibinfo  {journal} {Proceedings of the ACM on Human-Computer
  Interaction}\ }\textbf {\bibinfo {volume} {2}},\ \bibinfo {pages} {1}
  (\bibinfo {year} {2018})}\BibitemShut {NoStop}%
\bibitem [{\citenamefont {{Bill Snyder}}(2019)}]{AlexStamos}%
  \BibitemOpen
  \bibfield  {author} {\bibinfo {author} {\bibnamefont {{Bill Snyder}}},\
  }\href@noop {} {\enquote {\bibinfo {title} {{Alex Stamos}: How do we preserve
  free speech in the era of fake news?}}\ }\bibinfo {howpublished}
  {\url{https://engineering.stanford.edu/magazine/article/alex-stamos-how-do-we-preserve-free-speech-era-fake-news}}
  (\bibinfo {year} {2019})\BibitemShut {NoStop}%
\bibitem [{\citenamefont {Alfano}\ \emph {et~al.}(2020)\citenamefont {Alfano},
  \citenamefont {Fard}, \citenamefont {Carter}, \citenamefont {Clutton},\ and\
  \citenamefont {Klein}}]{alfano2020technologically}%
  \BibitemOpen
  \bibfield  {author} {\bibinfo {author} {\bibfnamefont {M.}~\bibnamefont
  {Alfano}}, \bibinfo {author} {\bibfnamefont {A.~E.}\ \bibnamefont {Fard}},
  \bibinfo {author} {\bibfnamefont {J.~A.}\ \bibnamefont {Carter}}, \bibinfo
  {author} {\bibfnamefont {P.}~\bibnamefont {Clutton}}, \ and\ \bibinfo
  {author} {\bibfnamefont {C.}~\bibnamefont {Klein}},\ }\href@noop {}
  {\bibfield  {journal} {\bibinfo  {journal} {Synthese}\ } (\bibinfo {year}
  {2020})}\BibitemShut {NoStop}%
\end{thebibliography}
\end{document}